\documentclass{article}
\pdfoutput=1

\usepackage{color}
\usepackage{graphicx,amssymb,amsfonts,amsmath,amssymb,amscd,amstext, mathrsfs}
\usepackage[colorlinks=true,linkcolor=blue,citecolor=black]{hyperref}

\def\be{\begin{eqnarray}}
\def\ee{\end{eqnarray}}

\def\csch{\operatorname{csch}}

\def\tr{\operatorname{tr}}

\def\Vol{\operatorname{Vol}}
\def\vol{\operatorname{vol}}

\def\tr{\operatorname{tr}}

\def\d{{\rm d}}

\newcommand{\D}{D}
\newcommand{\dW}{\mathcal{W}}

\newcommand{\Log}{\ln}
\newcommand{\beq}{\begin{equation}}
\newcommand{\eeq}{\end{equation}}
\newcommand{\nn}{\nonumber}

\newcommand{\II}{\mathrm{I}\hspace{-0.8pt}\mathrm{I}}

\newcommand{\tG}{\tilde{\Gamma}}
\newcommand{\tR}{\tilde{R}}
\newcommand{\oG}{\mathring{\Gamma}}
\newcommand{\oR}{\mathring{R}}
\newcommand{\oD}{\mathring{{\rm D}}}

\newcommand{\og}{\mathring{g}}

\newcommand{\W}{\mathcal{W}}
\newcommand{\oBox}{\mathring{\Box}}

\newcommand{\R}{\mathcal{R}}

\newcommand{\Q}{\mathcal{Q}}
\newcommand{\B}{\mathcal{B}}
\renewcommand{\D}{\mathcal{D}}

\begin{document}
\thispagestyle{empty}

\begin{flushright}
YITP-SB-15-37 
\end{flushright}


\begin{center}
\vspace{1.7cm} { \LARGE {\bf 
Universal Entanglement and Boundary Geometry \\ \vspace{.2cm} in Conformal Field Theory
}}
\vspace{1.1cm}

Christopher P. Herzog, Kuo-Wei Huang, and Kristan Jensen \\
\vspace{1.1 cm}
{ \it C.~N.~Yang Institute for Theoretical Physics,\\
Department of Physics and Astronomy, \\
Stony Brook University, Stony Brook, NY  11794}

\vspace{0.8cm}
\end{center}

\begin{abstract}
\noindent
Employing a conformal map to hyperbolic space cross a circle, we compute the universal contribution to the vacuum entanglement entropy (EE) across a sphere in even-dimensional conformal field theory. Previous attempts to derive the EE in this way were hindered by a lack of knowledge of the appropriate boundary terms in the trace anomaly. In this paper we show that the universal part of the EE can be treated as a purely boundary effect. As a byproduct of our computation, we derive an explicit form for the A-type anomaly contribution to the Wess-Zumino term for the trace anomaly, now including boundary terms. In $d=4$ and 6, these boundary terms generalize earlier bulk actions derived in the literature. 

\end{abstract}

\pagebreak
\setcounter{page}{1}

\tableofcontents

\section{Introduction}

Entanglement entropy has played an increasingly important role in theoretical physics. Invented as a measure of quantum entanglement, it has been successfully applied in a much broader context.  Entanglement entropy can serve as an order parameter for certain exotic phase transitions \cite{Osborne:2002zz,Vidal:2002rm}. It is likely very closely related to black hole entropy \cite{Bombelli:1986,Srednicki:1993im}. Certain types of entanglement entropy order quantum field theories under renormalization group flow~\cite{Casini:2006es,Casini:2012ei,Solodukhin:2008dh,Komargodski:2011vj}. It is the last result which is most relevant to this paper. In even space-time dimension, the connection between entanglement entropy and renormalization group flow is tied up in the existence of a Weyl, or trace, anomaly~\cite{Casini:2006es,Solodukhin:2008dh,Komargodski:2011vj}. In fact, certain universal terms in the entanglement entropy can be extracted from the anomaly. The moral of this paper is that to use the anomaly correctly, one should understand how to write it down on a manifold with a codimension one boundary.

To define entanglement entropy, we assume that the Hilbert space can be factorized, ${\mathcal H} = {\mathcal H}_A \otimes {\mathcal H}_B$, where ${\mathcal H}_A$ corresponds to the Hilbert space for a spatial region $A$ of the original quantum field theory.\footnote{This factorization is a nontrivial assumption. The boundary between $A$ and $B$, $\partial A$, plays an important role in recent discussons regarding the entanglement entropy of gauge theory~\cite{Buividovich:2008gq,Casini:2013rba,Donnelly:2014fua,Huang:2014pfa}. The boundary terms associated with $\partial A$ we find in this paper suggests that the factorization is not always a clean and unambiguous procedure even for non-gauge theories.} Given such a factorization one can construct the reduced density matrix $\rho_A = \tr_B \rho$ by tracing over the degrees of freedom in the complementary region $B$, where $\rho$ is the initial density matrix.  The entanglement entropy is the von Neumann entropy of the reduced density matrix:
\be
S_E \equiv - \tr (\rho_A \ln \rho_A) \ .
\ee  
Only when $\rho = |\psi\rangle \langle \psi |$ is constructed from a pure state $| \psi \rangle$ does $S_E$ measure the quantum entanglement.  Otherwise, it is contaminated by the mixedness of the density matrix $\rho$.

In a quantum field theory context, the definition of $S_E$ presents a challenge because the infinite number of short distance degrees of freedom render $S_E$ strongly UV divergent.  Consider for example a $d$-dimensional conformal field theory (CFT) in the vacuum.  Let $d$ be even so that the theory may have a Weyl anomaly, and let $A$ be a $(d-1)$-dimensional ball of radius $\ell$.  In this case, the entanglement entropy has an expansion in a short distance cut-off $\delta$
of the form
\be
S_E = \alpha \frac{\operatorname{Area}(\partial A)}{\delta^{d-2}} + \ldots + 4 a  (-1)^{d/2} \Log \frac{\delta}{\ell} + \ldots
\label{SEball}
\ee
The constant $\alpha$ multiplying the leading term is sensitive to the definition of the cut-off $\delta$ and thus has no physical meaning.  The fact that the leading term scales with the area of the boundary of $A$, however, is physical and suggests that most of the correlations in the vacuum are local.  

Most important for this paper, the subleading term in eq.~\eqref{SEball} proportional to the logarithm is ``$a$,'' the coefficient multiplying the Euler density in the trace anomaly~\cite{Deser:1993yx}
\be
\label{traceanomaly}
\langle {T^\mu}_\mu \rangle = \sum_j c_j I_j - (-1)^{d/2} \frac{4a}{d! \Vol(S^d)} E_d +
{\rm D}_{\mu}J^{\mu}  \,,
\ee
with ${\rm D}_{\mu}$ the covariant derivative. In this expression, $E_d$ is the Euler density normalized such that integrating $E_d$ over an $S^d$ yields $d! \Vol(S^d)$. See section \ref{sec:dgtr2} for more details about the definition of $E_d$. The $I_j$ are curvature invariants which transform covariantly with weight ${-}d$ under Weyl rescalings. There is also a total derivative ${\rm D}_\mu J^{\mu}$ whose precise form depends on the particular regularization scheme used in defining the partition function.\footnote{In the terminology of ref.~\cite{Deser:1993yx}, the Euler term is a type-A anomaly and the Weyl-covariants $I_j$ are type-B.}

The motivation for this paper is a puzzle described in ref.~\cite{Casini:2011kv}. The authors describe several different methods for verifying the logarithmic contribution to the entanglement entropy in~\eqref{SEball}. One is to conformally map the causal development of the ball, $\mathcal{D}$, to the static patch of de Sitter spacetime, and then exploit the trace anomaly~\eqref{traceanomaly}. Another method runs into difficulties. They attempt to compute $S_E$ by mapping $\mathcal{D}$ to hyperbolic space.  Here, the authors were not able to use the anomaly directly. Instead, they resorted to an effective anomaly action, which here fails because hyperbolic space has a boundary.  As we explain, and as was anticipated in ref.~\cite{Casini:2011kv}, getting the correct answer requires a careful treatment of boundary terms in the effective anomaly action.

To our knowledge, the relation between these boundary terms and entanglement entropy has not been considered carefully before.\footnote{In a somewhat different vein, there is a discussion of entanglement entropy on spaces with boundary in ref.~\cite{Fursaev:2013mxa}.} In $d=2$, the boundary contribution to the trace anomaly is textbook material~\cite{Polchinski}. In $d=4$ and $d=6$, the bulk anomaly induced dilaton effective actions are written down in refs.\ \cite{Komargodski:2011vj} and \cite{Elvang:2012st} respectively. (See also \cite{BrownOttewill} for $d=4$.)  Given the importance of the dilaton effective action in understanding the $a$-theorem \cite{Komargodski:2011vj}, and the recent ``$b$-theorem'' in $d=3$~\cite{Jensen:2015swa}, it seems conceivable the boundary correction terms may be useful in a more general context. In this paper we generalize these dilaton effective actions with boundary terms for a manifold with codimension one boundary and we show that these boundary terms are crucial in computing entanglement entropy.  We also provide a general procedure, valid in any even dimension, for computing these boundary terms.

We begin with the two-dimensional case in section \ref{sec:twod}, where we illustrate our program and use an anomaly action with boundary terms to recover the well-known results of the interval R\'enyi entropy~\cite{KorepinRenyi,Calabrese:2004eu} and the Schwarzian derivative. In section \ref{sec:dgtr2}, we construct the boundary terms in the trace anomaly in $d>2$ and present an abstract formula for the anomaly action in arbitrary even dimension.  We demonstrate the result satisfies Wess-Zumino consistency. In section \ref{sec:dilaton}, we compute the anomaly action in four and six dimensions, keeping careful track of the boundary terms. (In six dimensions, our boundary action is only valid in a conformally flat space time, while in four dimensions, the answer provided is completely general.) In section \ref{sec:example}, we resolve the puzzle of how to compute the entanglement entropy of the ball through a map to hyperbolic space in general dimension. The resolution of this puzzle constitutes the main result of the paper. We also revisit the computation of the entropy in de Sitter spacetime. Finally, we conclude in section \ref{sec:discussion}. We relegate various technical details to appendices. Appendix \ref{app:diffgeom} reviews some useful differential geometry for manifolds with boundary. Appendix \ref{A:WZ4d} contains a detailed check of Wess-Zumino consistency in four dimensions. Appendix \ref{app:dimreg} contains details of the derivation of the anomaly action in four and six dimensions. Appendix \ref{app:holo} provides a corresponding holographic calculation of entanglement entropy through a map to hyperbolic space.

\section{The Two Dimensional Case and R\'enyi entropy}
\label{sec:twod}

In two dimensions, the stress tensor has the well known trace anomaly
\be
\label{Tmunu2d}
\langle {T^\mu}_\mu \rangle = \frac{c}{24 \pi} R\,,
\ee
where we have replaced the anomaly coefficient $a$ with the more common central charge $c = 12 a$ which appears in the two-point correlation function of the stress tensor. Eq.\ \eqref{Tmunu2d} is the Ward identity for the anomalous Weyl symmetry. It is equivalent to the variation of the generating functional $W[g_{\mu\nu}] = - \Log Z [g_{\mu\nu}]$ under a Weyl variation $\delta g_{\mu\nu} = 2  g_{\mu\nu}\delta \sigma $. However, on a manifold with boundary, the anomalous variation of $W$ may contain a boundary term. In this section, we show how to construct the anomaly effective action with boundary terms for the simplest case, $d=2$. We will reproduce the classic entanglement entropy result using the boundary term in the anomaly action. We also show that the boundary term correctly recovers the universal term in the single-interval $d=2$ R\'enyi entropy.

\subsection{Anomaly Action with Boundary and Entanglement Entropy} In $d=2$, the most general result for the Weyl variation of the partition function consistent with Wess-Zumino consistency is
\cite{Polchinski}
\be
\label{E:deltaW}
\delta_{\sigma} W =- \frac{c}{24 \pi} \left[ \int_M \d^2 x \sqrt{g} R \, \delta \sigma + 2 \int_{\partial M} \d y \sqrt{\gamma} K \, \delta \sigma \right] \ .
\ee
To write this expression, we have introduced some notation.  In $d=2$, the notation is overkill, but we need the full story in what follows in $d>2$. We denote bulk coordinates as $x^{\mu}$ and boundary coordinates as $y^{\alpha}$. Let $n^\mu$ be the unit-length, outward pointing normal vector to $\partial M$ and $\gamma_{\alpha\beta}$ the induced metric on $\partial M$.  
We can define $K$ in two equivalent ways. First, locally near the boundary we can extend $n^{\mu}$ into the bulk. We can choose to extend it in such a way that $n^\mu {\rm D}_\mu n_\nu = 0$,  
in which case the extrinsic curvature is defined to be $K_{\mu\nu} \equiv {\rm D}_{(\mu} n_{\nu)}$. The trace of the extrinsic curvature is $K = {K^\mu}_\mu$. Alternatively, we can also define $K$ purely from data on the boundary. The bulk covariant derivative ${\rm D}_{\mu}$ induces a covariant derivative $\oD_{\alpha}$ on the boundary. It can act on tensors with bulk indices, boundary indices, or mixed tensors with both. We specify the boundary through a map $\partial M \to M$, which amounts to a set of $d$ embedding functions $X^{\mu}(y^{\alpha})$. The $\partial_{\alpha} X^{\mu}$ are tensors on the boundary, and their derivative gives the extrinsic curvature as $K_{\alpha\beta} = - n_{\mu}\oD_{\alpha} \partial_{\beta} X^{\mu}$, and its trace $K = \gamma^{\alpha\beta}K_{\alpha\beta}$. For more details on differential geometry of manifolds with boundary, see appendix \ref{app:diffgeom}.

Observe that, for a constant Weyl rescaling $\delta \sigma = \lambda$, the Weyl anomaly~\eqref{E:deltaW} is equivalent to
\be
\delta_{\lambda} W = - \frac{c}{6} \chi \lambda \,,
\ee
where $\chi$ is the Euler characteristic of $M$. That is, the boundary term in the Weyl anomaly is simply the boundary term in the Euler characteristic.

The stress tensor is defined as
\be
\langle T^{\mu\nu}\rangle = -\frac{2}{\sqrt{g}}\frac{\delta W}{\delta g_{\mu\nu}}\,,
\ee
in which case~\eqref{E:deltaW} leads to a boundary term in the trace of the stress tensor,
\be
\langle T^{\mu}{}_{\mu}\rangle = \frac{c}{24 \pi}\left( R + 2 K \delta(x^{\perp})\right)\,,
\ee
where $\delta(x^{\perp})$ is a Dirac delta function with support on the boundary.

We now wish to write down a local functional which reproduces the variation~\eqref{E:deltaW}. To do so we introduce an auxiliary ``dilaton'' field $\tau$ which transforms under a Weyl transformation $g_{\mu\nu} \to e^{2 \sigma} g_{\mu\nu}$ as $\tau \to \tau + \sigma$. The quantity 
\be
\hat{g}_{\mu\nu} \equiv e^{-2 \tau} g_{\mu\nu}\,,
\ee
is invariant under this generalized Weyl scaling and so too the effective action $\hat{W} \equiv W[e^{-2 \tau} g_{\mu\nu}]=W[\hat{g}_{\mu\nu}]$.
Then
\be
\dW[g_{\mu\nu}, e^{-2 \tau} g_{\mu\nu} ] \equiv   W - \hat{W}\,,
\ee
will vary to reproduce the anomaly, $\delta_{\sigma} \dW = \delta_{\sigma}  W$. 

In what follows, we refer to $\dW$ as a ``dilaton effective action,'' given its similarities with the dilaton effective action presented in refs.\ \cite{Komargodski:2011vj,Elvang:2012st}. However, unlike those works we are only considering conformal fixed points and not renormalization group flows, and so this name is a bit of a misnomer. More precisely, $\dW$ is a Wess-Zumino term for the Weyl anomaly, or alternatively an anomaly effective action. Analytically continuing to Lorentzian signature, it computes the phase picked up by the partition function under the Weyl rescaling from a metric $g_{\mu\nu}$ to $e^{-2\tau}g_{\mu\nu}$.

What exactly is $\dW$ in $d=2$?  The first quick guess is
\be
\dW_0 = -\frac{c}{24 \pi} \left[ \int_M \d^2 x \sqrt{g} R \tau + 2 \int_{\partial M} \d y \sqrt{\gamma} K \tau \right] \ .
\ee 
But the metric scales, and we should take into account that under Weyl scaling in $d=2$,
\begin{align}
\begin{split}
R[e^{2 \sigma} g_{\mu\nu}] &= e^{-2 \sigma} ( R[g_{\mu\nu}] - 2 \Box \sigma ) \,,
\\
K[e^{2 \sigma} g_{\mu\nu}] &= e^{-\sigma} (K[g_{\mu\nu}] + n^\mu \partial_\mu \sigma ) \, .
\end{split}
\end{align}
To cancel these variations, we add a $(\partial \tau)^2 \equiv (\partial_\mu \tau) (\partial^\mu \tau)$ term to the effective action. The total effective anomaly action is then
\be
\dW = - \frac{c}{24 \pi} \left[ \int_M \d^2 x \sqrt{g} \left( R[g_{\mu\nu}] \tau - (\partial \tau)^2 \right) + 2 \int_{\partial M} \d y \sqrt{\gamma} \, K[g_{\mu\nu}] \tau \right] + (\text{invariant}) \ .
\label{E:almostThere2d}
\ee
The right-hand side is computed with the original unscaled metric $g_{\mu\nu}$.\footnote{
 This action corrects a typo in eq. (1.2) of ref.\ \cite{Komargodski:2011xv}, as well as accounts for the boundary term.}
In writing~\eqref{E:almostThere2d}, we have allowed for the possibility of additional terms invariant under the Weyl symmetry. There are only two such terms with dimensionless coefficients,
\be
\int_M \d^2x \sqrt{\hat{g}}\hat{R}\,, \qquad \int_{\partial M} \d y\sqrt{\hat{\gamma} }\hat{K}\,.
\ee
However, now we use that by definition $\dW=0$ when $\tau=0$. Thus neither of these terms can appear in $\dW$, so
\be
\dW = - \frac{c}{24 \pi} \left[ \int_M \d^2 x \sqrt{g} \left( R[g_{\mu\nu}] \tau - (\partial \tau)^2 \right) + 2 \int_{\partial M} \d y \sqrt{\gamma} \, K[g_{\mu\nu}] \tau \right] \ .
\label{Wefftwod}
\ee

The second step, which involved adding by hand a $(\partial \tau)^2$ term to cancel some unwanted pieces of the Weyl variation, seemed to involve some guess work which could become a problem in $d>2$ where the expressions are much more complicated.  In fact, there are several constructive algorithms which remove this element of guesswork.  
One method involves integrating the anomaly \cite{Wess:1971yu, Cappelli:1989, Schwimmer:2010za}:
\begin{align}
\begin{split}
\label{integratedanomaly}
\dW &=- \frac{c}{24 \pi} \int_0^1 \d t \left. \left[ \int_M \d^2 x \,  \sqrt{g'} \,  R[ g'_{\mu\nu}] \tau
+ 2 \int_{\partial M} \d y \,  \sqrt{\gamma'} \, K[ g'_{\mu\nu}] \tau \right] \right|_{g'_{\mu\nu} = e^{-2 t \tau} g_{\mu\nu}} \,
\\
& = - \int_0^1dt \left.\int_M \d^2x \sqrt{g'}\, \langle T^{\mu}{}_{\mu}[g'_{\nu\rho}]\rangle \tau\right|_{g'_{\mu\nu}=e^{-2t\tau}g_{\mu\nu}}\,.
\end{split}
\end{align}
Thus, given the trace anomaly $\langle {T^\mu}_\mu \rangle$, it is straightforward albeit messy to reconstruct $\dW$.

The second method (which we elaborate in this paper) is dimensional regularization \cite{Brown:1977sj,Christensen:1978sc}.  We define $\widetilde W[g_{\mu\nu}]$ in $n = 2 + \epsilon$ dimensions:
\be
\widetilde W[g_{\mu\nu}] \equiv -\frac{c}{24 \pi (n-2)} \left[ \int_M \d^n x \, \sqrt{g} \, R + 2\int_{\partial M} \d^{n-1} y \, \sqrt{\gamma} \, K \right] \ ,
\label{tildeWtwo}
\ee
where $R$, $K$, $g_{\mu\nu}$, and $\gamma_{\alpha\beta}$ are dimensionally continued in the naive way. We claim then that
\be
\dW = \lim_{n \to 2} \left( \widetilde W[g_{\mu\nu}] - \widetilde W[e^{-2 \tau} g_{\mu\nu}] \right) \,,
\ee
as one may verify after a short calculation, using the more general rules for the Weyl transformations in $n$ dimensions,
\begin{align}
\begin{split}
R[e^{2 \sigma} g_{\mu\nu}] &= e^{-2 \sigma} \left( R[g_{\mu\nu}] - 2 (n-1) \Box \sigma - (n-2)(n-1) (\partial \sigma)^2 \right) \,, \\
K[e^{2 \sigma} g_{\mu\nu}] &= e^{-\sigma} \left( K[g_{\mu\nu}] + (n-1) n^\mu \partial_\mu \sigma \right) \,.
\end{split}
\end{align}

In all three cases, we are computing the same difference between two effective actions. It would be preferable to have access to the effective actions themselves.  There are two problems here. The full actions depend on more than the anomaly coefficients.  They are also likely to be ultraviolet and perhaps also infrared divergent. If we focus just on the anomaly dependent portion, it could easily be that some of this anomaly dependence is invariant under Weyl scaling and drops out of the difference we have computed.
Interestingly, the dimensional regularization procedure offers a regulated candidate $\widetilde W[g_{\mu\nu}]$ for the anomaly dependent portion of $W[g_{\mu\nu}]$.  

Let us try to extract some information from the regulated candidate action in flat space:
\be
\label{2dflat}
\widetilde W[\delta_{\mu\nu}] =- \frac{c}{12 \pi (n-2)} \int_{\partial M} \d^{n-1} y \sqrt{\gamma} K \ .
\ee
A simple case, which also turns out to be relevant for the entanglement entropy calculations we would like to perform, is where $M$ is a large ball of radius $\Lambda$ with a set of $q$ smaller, non-intersecting balls of radius $\delta_j$ removed. For each ball, we can work in a local coordinate system where $r$ is a radial coordinate. For the smaller balls, $\sqrt{\gamma} K = -r^{n-2}$ while for the large ball $\sqrt{\gamma} K = r^{n-2}$. It then follows that 
\be
\label{Wflat}
\widetilde W[\delta_{\mu\nu}] = -\frac{c}{6} \left[ \frac{1}{n-2} (1-q) + \frac{q+1}{2}(\gamma + \Log \pi) + \Log \Lambda - \sum_{j=1}^{q} \Log \delta_j + O(n-2) \right] \ .
\ee
The leading divergent contribution is proportional to the Euler characteristic $\chi = 1-q$ of the surface.

We claim that the $\ln \delta_j$ pieces of the expression~\eqref{Wflat} can be used to identify a universal contribution to the entanglement entropy of a single interval in flat space.  We will justify the computation through a conformal map to the cylinder, but in brief, the computation goes as follows. For an interval on the line with left endpoint $u$ and right endpoint $v$, to regulate the UV divergences in the entanglement entropy computation
we place small disks around the points $u$ and $v$ with radius $\delta$.  The entanglement entropy then turns out to be the logarithmic contribution of these disks to $-\widetilde W[\delta_{\mu\nu}]$:
\be
S_E \sim -\frac{c}{3} \Log \delta \ .
\ee
As the underlying theory is conformal, the answer can only depend on the conformal cross ratio of the two circles $4 \delta^2 / |u-v|^2$.  Thus we find the classic result \cite{Holzhey:1994we,Calabrese:2004eu}
\be
\label{ee}
S_E \sim \frac{c}{3} \Log \frac{|v-u|}{\delta} \ . 
\ee
Here and henceforth, the $\sim$ indicates that the LHS has a logarithmic dependence given by the RHS. We neglect the computation of the constant quantity in $S_E$, as it depends on the precise choice of regulator and so is unphysical.

A more thorough justification of this computation occupies the next two subsections.  In broad terms, the same result turns out to be valid in even dimensions $d>2$, a fact whose demonstration will occupy most of the remainder of the paper.  More specifically, we mean that the logarithmic contribution to $\widetilde W[\delta_{\mu\nu}]$ for flat space with $D \times S^{d-2}$ removed, where $D$ is a small disk of radius $\delta$, yields a universal contribution to entanglement entropy for a ball shaped region in flat space.

To return to $d=2$, we describe the plane to the cylinder map and its relevance for entanglement entropy 
in section \ref{sec:planetocylinder}.  The demonstration however requires we also know how the stress tensor transforms under conformal transformations.  The transformation involves the Schwarzian derivative which can be found in most textbooks on conformal field theory.  In an effort to be self contained we will use our effective anomaly action to derive the Schwarzian derivative
in section \ref{sec:schwarzian}.  
In $d=2$, the effective action turns out to be useful to compute not only the entanglement entropy but also the single interval 
R\'enyi entropies.
A calculation of the R\'enyi entropies is provided in section \ref{sec:renyi}.

\subsection{The Schwarzian Derivative}
\label{sec:schwarzian}

To calculate the change in the stress tensor under a Weyl scaling from $g_{\mu\nu}$ to $\hat{g}_{\mu\nu}=e^{-2\tau}g_{\mu\nu}$, we begin with a variation of $\dW = W-\hat{W}$ with respect to the metric $g_{\mu\nu}$,
\begin{align}
\begin{split}
\label{E:transOfT}
\delta \dW = & \delta W - \delta \hat{W}
\\
= & - \frac{1}{2}\int_M \d^2x \left( \sqrt{g}\,\delta g_{\mu\nu}\langle T^{\mu\nu}\rangle_g - \sqrt{\hat{g}} \,\delta \hat{g}_{\mu\nu}\langle T^{\mu\nu}\rangle_{\hat{g}}\right)
\\
= & - \frac{1}{2}\int_M \d^2x \sqrt{g} \,\delta g_{\mu\nu}\left( \langle T^{\mu\nu}\rangle_g - e^{-4 \tau} \langle T^{\mu\nu}\rangle_{\hat{g}}\right)\,,
\end{split}
\end{align}
where in the last line we have used that $\sqrt{\hat{g}}\delta \hat{g}_{\mu\nu} = \sqrt{g}\,e^{-(d+2)\tau}\delta g_{\mu\nu}$ in $d$ dimensions. The subscript $g$ on the expectation value refers to $\langle T^{\mu\nu}\rangle$ on the space with metric $g$, and similarly for $\hat{g}$. Using the explicit expression for $\dW$ in~\eqref{Wefftwod}, we compute its variation
\begin{align}
\begin{split}
\label{E:deltaWZ2d}
\delta \dW &=
-\frac{c}{24\pi}\int_M \d^2x \sqrt{g} \,\delta g_{\mu\nu}\left[ \partial^{\mu}\tau\partial^{\nu}\tau+{\rm D}^{\mu}{{\partial}}^{\nu}\tau-g^{\mu\nu}\left( \frac{1}{2}(\partial\tau)^2 +\Box\tau\right)\right] 
\\
& \qquad -\frac{c}{24\pi}\int_{\partial M}\d y \sqrt{\gamma}\,\delta g_{\mu\nu} h^{\mu\nu} n^{\rho}\partial_{\rho}\tau\,,
\end{split}
\end{align}
where $h^{\mu\nu}$ is the projector to the boundary,
\beq
h_{\mu\nu}=g_{\mu\nu}-n_{\mu}n_{\nu}\,.
\eeq
In obtaining~\eqref{E:deltaWZ2d} we have used that in two dimensions the Einstein tensor $R_{\mu\nu}-\frac{R}{2}g_{\mu\nu}$ vanishes, and that the variation of the Ricci tensor is a covariant derivative
$\delta R_{\mu\nu} = {\rm D}_{\rho}\delta \Gamma^{\rho}{}_{\mu\nu}-{\rm D}_{\nu}\delta\Gamma^{\rho}{}_{\mu\rho}$. Putting~\eqref{E:deltaWZ2d} together with~\eqref{E:transOfT}, we find
\be
\langle T_{\mu\nu}\rangle_{\hat{g}} = \langle T_{\mu\nu}\rangle_g - \frac{c}{12\pi}\left[ \partial_{\mu}\tau\partial_{\nu}\tau + {\rm D}_{\mu}{\partial}_{\nu}\tau - g_{\mu\nu}\left( \frac{1}{2}(\partial\tau)^2 + \Box \tau\right)\right] - \frac{c}{12\pi}\delta(x^{\perp}) h_{\mu\nu}n^{\rho}\partial_{\rho}\tau \,.
\ee

Suppose we consider a Weyl rescaling which takes us from flat space, $g_{\mu\nu}=\delta_{\mu\nu}$, to the new metric $\hat{g}_{\mu\nu}=e^{-2\tau}\delta_{\mu\nu}$. The stress tensor for a conformal theory in vacuum on the plane is usually defined to vanish. Thus the stress tensor on the manifold with metric $e^{-2 \tau} \delta_{\mu\nu}$ will be
\beq
\langle T_{\mu\nu} \rangle = -\frac{c}{12 \pi} \left[ \partial_\mu \tau \partial_\nu \tau+  \partial_\mu \partial_\nu \tau
- \delta_{\mu\nu} \left( \frac{1}{2} (\partial \tau)^2 + (\Box \tau) \right) \right] \ 
\eeq
(dropping the boundary contribution).
The Schwarzian derivative describes how the stress tensor transforms under a conformal transformation, i.e.\ a combination of a Weyl rescaling and a diffeomorphism that leaves the metric invariant. If the complex plane is parametrized initially by $z$ and $\bar z$, we introduce new variables $w(z)$ and $\bar w(\bar z)$ and require that the Weyl rescaling satisfies
\beq
e^{-2 \tau} = \left( \frac{\partial w}{\partial z} \right) \left( \frac{ \partial \bar w}{\partial \bar z} \right)  \ .
\eeq
Start with the stress tensor in the $w$-plane, and perform a diffeomorphism to go to the $z$ variables.  That transformed stress tensor should be related by a Weyl rescaling by $e^{-2\tau}$ to the stress tensor on the flat complex $z$-plane. Recalling that $g_{zz} = 0$, we find that
\be
(\partial_z w)^2 \langle T_{ww}(w) \rangle = \langle T_{zz}(z) \rangle_{e^{-2 \tau} \delta_{\mu\nu}} &=& -\frac{c}{12 \pi} \left[ (\partial_z \tau)^2 + (\partial_z^2 \tau) \right] \nonumber \\
&=& \frac{c}{48 \pi} \frac{2 (\partial_z^3 w)(\partial_z w) - 3 (\partial_z^2 w)^2}{ (\partial_z w)^2} \ ,
\label{schwarzian}
\ee
which is the usual result for the Schwarzian derivative.

\subsection{Entanglement Entropy from the Plane and Cylinder}
\label{sec:planetocylinder}

\begin{figure}
\begin{center}
\includegraphics[width=2.5in]{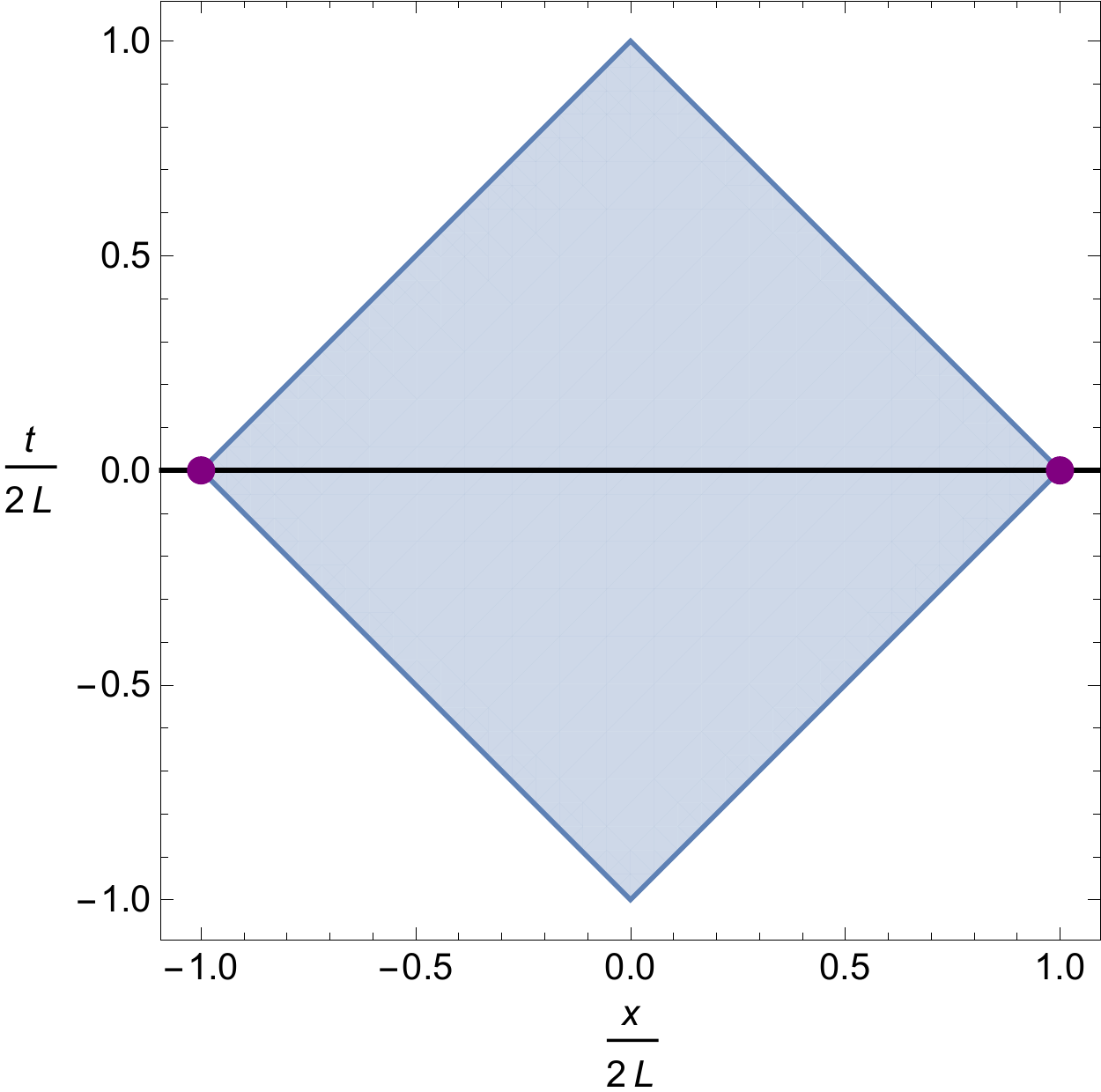}
\caption{\label{F:causal2d} The causal development of an interval of length $L$. The dots indicate the endpoints of the interval.}
\end{center}
\end{figure}

We now consider the entanglement entropy of an interval with left endpoint $u$ and right endpoint $v$. The information necessary to compute the entropy is contained in the causal development of this interval, i.e.\ the diamond shaped region bounded by the four null lines $x \pm t = u$ and $x \pm t = v$. See Fig.~\ref{F:causal2d}.
%
We will indirectly deduce the entanglement entropy by conformally mapping to a thermal cylinder, keeping careful track of the phase picked up by the partition function under the transformation. 

Consider the following change of variables
\be
e^{2 \pi w / \beta} = \frac{z-u}{z-v} \ ,
\label{ztow}
\ee
where $z = x - t = x + i t_{\rm E}$, and correspondingly for $\bar z$ and $\bar w$.  If we let $w = \sigma^1 + i \sigma^2$, then $\sigma^2$ is periodic with periodicity $\beta$, $\sigma^2 \sim \sigma^2 + \beta$. In other words, the theory on the $w$-plane is naturally endowed with a 
 temperature $1/\beta$. The other nice property of this map is that the 
the interval at time $t=0$ is mapped to the real line $\mbox{Re}(w)$. Thus the reduced density matrix $\rho_A$ associated with the interval is related by a unitary transformation to the thermal density matrix $\rho_\beta$ on the line. As the entanglement entropy is invariant under unitary transformations, the entanglement entropy of the interval is the thermal entropy associated with the cylinder, that is the thermal entropy on the infinite line. If we let
\be
\rho = \frac{e^{-\beta H}}{\tr e^{-\beta H}}  \ ,
\ee
where $H$ is the Hamiltonian governing evolution on the line, then
\be
S_E = -\tr (\rho \ln \rho)= \beta \tr (\rho H)+\ln \tr (e^{-\beta H})= \beta \langle H \rangle - W_{\rm cyl} \ ,
\ee
where $W_{\rm cyl} \equiv -\ln \tr e^{-\beta H}$ is the partition function on the cylinder. This entropy is infinite because the cylinder is infinitely long in the $\sigma^1$ direction, and we need to regulate the divergence. The natural way to regulate is to cut off the cylinder such that $-\Lambda < \sigma^1 < \Lambda$. In the $z = x + i t_{\rm E}$ plane, these cut-offs correspond to putting small disks of radius $\delta$ around the endpoints $u$ and $v$, where now
\be
\frac{\delta}{v-u} = e^{-2 \pi \Lambda/\beta} \ .
\ee

We have two quantities to compute, $\beta \langle H \rangle$ and $W_{\rm cyl}$.  We can use the Schwarzian derivative from the previous subsection to compute 
\be
\beta \langle H \rangle = \int_{\rm cyl} \langle T^{00}\rangle \d \sigma^1 \, ,
\ee
where we have analytically continued $\sigma^0=-i \sigma^2$. From the transformation rules~\eqref{schwarzian} and~\eqref{ztow}, the $ww$ component of the stress tensor on the cylinder is
\be
\langle T_{ww}(w)\rangle &=& \frac{\pi c}{12 \beta^2} \ .
\ee  
In Cartesian coordinates, $T^{22} = - \frac{1}{4} (T^{ww} + T^{\bar w \bar w})$.  Thus we have,
analytically continuing to real time $\sigma^0 = -i \sigma^2$, a positive thermal energy $\langle T^{00} \rangle = \frac{\pi c}{6 \beta^2} $ from which follows the first quantity of interest
\be
\beta \langle H \rangle = \frac{\pi c}{3 \beta} \Lambda  = \frac{c}{6} \Log \frac{|v-u|}{\delta}\ .
\ee

Toward the goal of computing $W_{\rm cyl}$, we first compute the difference in anomaly actions 
$\dW[\delta_{\mu\nu}, e^{-2 \tau} \delta_{\mu\nu}]$ where the dilaton $\tau$ is derived from the plane to cylinder map
\be
\tau = -\frac{1}{2} \ln \left[ \frac{\beta}{2\pi} \left( \frac{1}{v-z} - \frac{1}{u-z}\right) \right] + c.c. 
\ee
Given the dilaton, we can compute the bulk contribution to the difference in effective actions
\be
\int  \d^2x\sqrt{g}\,(\partial \tau)^2  =\left( \frac{\pi}{\beta} \right)^2 \int_{\rm cyl} \d w \, \d \bar w \left| \coth \frac{\pi w}{\beta} \right|^2
= \frac{8 \pi^2 }{\beta} \Lambda\ , 
\ee
and the boundary contribution
\be
-2\int \d y \sqrt{\gamma} \,K \tau \sim  8 \pi \Log \delta  \sim -\frac{16 \pi^2}{\beta}  \Lambda \ .
\ee
Assembling the pieces, the difference in anomaly actions is then
\be
\dW[\delta_{\mu\nu}, e^{-2 \tau} \delta_{\mu\nu}] \sim- \frac{\pi c}{3 \beta} \Lambda = -\frac{c}{6} \Log \frac{|v-u|}{\delta} \ .
\ee
The last component we need is the universal contribution to $W[\delta_{\mu\nu}]$, which we claimed was actually equal to the universal contribution to single interval entanglement entropy.  Indeed, everything works as claimed since the contributions from $\beta \langle H \rangle$ and $\dW[\delta_{\mu\nu}, e^{-2 \tau} \delta_{\mu\nu}]$ cancel out:
\be
S_E = \beta \langle H \rangle + \dW[\delta_{\mu\nu}, e^{-2 \tau} \delta_{\mu\nu}]  - W[\delta_{\mu\nu}] \sim -\widetilde W[\delta_{\mu\nu}] \sim  \frac{c}{3} \Log \frac{|v-u|}{\delta} \ .
\ee

\subsection{R\'enyi Entropies from the Annulus}
\label{sec:renyi}

In $d=2$, the anomaly effective action also allows us to compute the R\'enyi entropies of an interval $A$,
\be
S_n \equiv \frac{1}{1-n} \Log \tr \rho_A^n \ .
\label{E:defRenyi}
\ee
We use the replica trick to compute $S_n$.  We can replace $\tr \rho_A^n$ with a certain ratio of Euclidean partition functions
\be
\tr \rho_A^n = \frac{Z(n)}{Z(1)^n} \ ,
\ee
where $Z(n)$ is the path integral on an $n$-sheeted cover of flat space, branched over the interval $A$. In the present case, we can use the coordinate transformation,
\be
w = \frac{z-u}{z-v}  \ ,
\ee
to put the point $u$ at the origin and the point $v$ at infinity.  As is familiar from the computation in the previous subsection, we need to excise small disks around the points $u$ and $v$, or correspondingly restrict to an annulus in the $w$ plane of radius $r_{\rm min} < r < r_{\rm max}$.  

To get the R\'enyi entropies, we would like to compare the partition function on the annulus to an $n$-sheeted cover of the annulus.  In two dimensions, these two metrics are related by a Weyl transformation. We take the metric on the annulus to be 
\be
g= \d r^2 + r^2 \d \theta^2 \ ,
\ee
while on the $n$-sheeted cover we have
\be
\hat{g} = e^{-2 \tau} g = \d \rho^2 + n^2 \rho^2 \d \theta^2 \ ,
\ee
 with $e^{-\tau} = n r^{n-1}$ and $\rho = r^n$. 
With this choice of $\tau$, the difference in anomaly actions becomes
\begin{align}
\begin{split}
\dW[\delta_{\mu\nu},e^{-2\tau} \delta_{\mu\nu}, ] &= \frac{c}{12} \left[ \int_{r_{\rm min}}^{r_{\rm max}} (\partial \tau)^2 r \, \d r - 2 \tau |_{r_{\rm min}}^{r_{\rm max}} \right]
\\
&= \frac{c}{12} (n^2 - 1) \Log \frac{r_{\rm max}}{r_{\rm min}} \ .
\end{split}
\end{align}
Now to isolate the universal contribution to $W[e^{-2 \tau} \delta_{\mu\nu}]$, we should remove the universal contribution from $W[\delta_{\mu\nu}]$:
\be
W[e^{-2 \tau} \delta_{\mu\nu}] \sim -\frac{c}{12} (n^2 + 1) \Log \frac{r_{\rm max}}{r_{\rm min}}\sim - \frac{c}{12}\left( n+\frac{1}{n}\right) \Log \frac{\rho_{\rm max}}{\rho_{\rm min}} \ .
\ee
We can tentatively identity this quantity with $-\Log Z(n)$.  To compute the R\'enyi entropies, we need to subtract off $n \Log Z(1)$. There is an issue here, however: both $\Log Z(n)$ and $\Log Z(1)$ are divergent quantities, and in comparing them we must arrange for the cutoffs to be congruous. We claim that in order to compare $\Log Z(n)$ with $\Log Z(1)$ we ought to use the $\rho$-cutoffs so that we excise discs of the same radius in each case. Thus, we need to subtract $n W[\delta_{\mu\nu}]$ using the cut-offs in the $\rho$ coordinate system, 
\be
\Log Z(n) - n \Log Z(1) \sim \frac{c}{12}\left( -n + \frac{1}{n} \right) \Log \frac{\rho_{\rm max}}{\rho_{\rm min}} \ .
\ee
Using the definition~\eqref{E:defRenyi} of the R\'enyi entropy, we find that
\be
S_n \sim \frac{c}{12} \left( 1 + \frac{1}{n}\right) \Log \frac{\rho_{\rm max}}{\rho_{\rm min}} \ .
\ee
Translating back to the $z$ plane, this result recovers the classic result \cite{KorepinRenyi, Calabrese:2004eu}\footnote{%
 The calculation we have just presented is very similar in spirit if not in detail to ones in refs.\ \cite{Solodukhin:1994yz,Frolov:1996hd}.
}
\be
S_n \sim \frac{c}{6} \left( 1 + \frac{1}{n}\right)\Log \frac{|v-u|}{\delta} \ .
\ee 
Taking $n\to 1$, it reduces to the previous entanglement entropy result \eqref{ee}. Note that in $d>2$, one still has an $n$-sheeted cover of an annulus, but it is less clear what to do with the remaining $d-2$ dimensions.

\section{Anomaly Actions in More than Two Dimensions}
\label{sec:dgtr2}

The trace anomaly~\eqref{traceanomaly} and effective anomaly action $\dW$ have an increasingly complicated structure as the dimension increases. Several issues need to be addressed for a complete treatment of the effective action. Before embarking, we warn the reader that this section is technical. The chief results are 1) the boundary term in the $a$-type anomaly~\eqref{Qdef} and~\eqref{E:aAnomalyWithBdy}, 2) two equivalent forms for the $a$-type anomaly action in~\eqref{E:integratedAnomalyGeneral} and~\eqref{tildeWgeneral}, and 3) a demonstration that the $a$-type anomaly, including the boundary term we obtain, is Wess-Zumino consistent in any dimension in subsection~\ref{S:WZconsistency}. Finally, 4) in~\eqref{E:4dTotalAnomaly} we present the most general form of the trace anomaly in $d=4$, including boundary central charges.

\subsection{Boundary Term of the Euler Characteristic}

As this paper was motivated by the problem of universal contributions to the entanglement entropy across a sphere in flat space, our main focus is on how the $a$ contribution to the anomaly action is modified in the presence of a boundary.  Regarding the other issues, we make a few preliminary comments which will be developed minimally in the rest of the paper.

The presence of a boundary affects the $c_j$ contributions to the trace anomaly~\eqref{traceanomaly} trivially.  Let us dispose of this issue immediately. The $I_j$ are, by definition, covariant under Weyl scaling. In fact the $\sqrt{g} I_j$ are invariant under Weyl scaling and so the $c_j$ contributions to $\dW[g_{\mu\nu}, e^{-2 \tau} g_{\mu\nu}]$ are simply
\be
\dW_c \equiv - \sum_j c_j \int_M \d^d x\, \sqrt{g} \, \tau I_j \ ,
\ee
with no additional boundary term.

The total derivative term in the trace anomaly \eqref{traceanomaly} 
depends on the choice of scheme. As we focus on universal aspects of the trace anomaly, with some exceptions we shall largely ignore this object in what follows. A fourth issue we have little to say about, with one exception, is the possible existence of additional terms in the trace anomaly associated purely with the boundary. These additional terms are best understood when the bulk CFT is odd-dimensional, so that the trace anomaly only has boundary terms. Those boundary terms can include the boundary Euler density as well as Weyl-covariant scalars~\cite{Graham:1999pm,Schwimmer:2008yh}, in analogy with the trace anomaly of even-dimensional CFT. See ref.~\cite{Jensen:2015swa}, which argued for a boundary ``$c$-theorem'' using this boundary anomaly. 
In this work we focus on CFTs in even dimension, with an odd-dimensional boundary. In $d=4$, using Wess-Zumino consistency, we identify two allowed boundary terms in the trace anomaly, but have nothing to add in $d \geq 6$.  

To return to the $a$-type anomaly, the central observation is that the $a$ dependent contribution to the trace anomaly~\eqref{traceanomaly} integrates to give a quantity proportional to the Euler characteristic for a manifold without boundary. The natural guess is then that in the presence of a boundary, one should add whatever boundary term is needed such that the integral continues to give a quantity proportional to the Euler characteristic. (Indeed we saw precisely this story play out in two dimensions in section \ref{sec:twod}.) 
The requisite boundary term is well known in the mathematics literature.  See for example the review \cite{Eguchi:1980jx}.  It is a Chern-Simons like term constructed from the Riemann and extrinsic curvatures. To write it down, we need some notation.      

We start by introducing the orthonormal (co)frame one forms $e^A = e^A_\mu dx^\mu$, in terms of which the metric on $M$ is $g_{\mu\nu}=\delta_{AB}e^A_{\mu}e^B_{\nu}$. Here and there, we also need their inverse $E^{\mu}_A$, satisfying $E^{\mu}_A e^A_{\nu}=\delta^{\mu}_{\nu}$ and $E^{\mu}_A e^B_{\mu} = \delta^A_B$. From the $e^A$ and the Levi-Civita connection $\Gamma^{\mu}{}_{\nu\rho}$, we construct the connection one-form ${\omega^A}_B$ via
\be
\partial_{\mu}e_{\nu}^A - \Gamma^{\rho}{}_{\nu\mu}e^A_{\rho} + \omega^A{}_{B\mu}e^B_{\nu}=0\,.
\label{oneformdef}
\ee
From this definition, it follows that $\omega^{AB}=-\omega^{BA}$ and the torsion one-form vanishes,
\be
\d e^A + {\omega^A}_B \wedge e^B = 0 \,.
\ee
Further, the curvature two-form built from $\omega^A{}_B$,
\be
\R^A{}_B \equiv \d\omega^A{}_B + \omega^A{}_C \wedge \omega^C{}_B  =\frac{1}{2}\mathcal{R}^A{}_{B\mu\nu}dx^{\mu}\wedge dx^{\nu}\,,
\ee
is related to the Riemann curvature by
\be
E^{\mu}_A \mathcal{R}^A{}_{B\rho\sigma} e^B_{\nu} = R^{\mu}{}_{\nu\rho\sigma}\,.
\ee
The curvature two-form satisfies the Bianchi identity
\be
\d\R^A{}_B + \omega^A{}_C \wedge \R^{CB} - \R^A{}_C \wedge \omega^C{}_B= 0 \ .
\ee
The Euler form is then
\be
{\mathcal E}_d \equiv \R^{A_1 A_2} \wedge \cdots \wedge \R^{A_{d-1} A_d}  \epsilon_{A_1 \cdots A_d} \ .
\label{E:defEuler}
\ee
where $\epsilon_{A_1 \cdots A_d}$ is the totally antisymmetric Levi-Civita tensor in dimension $d$. The Euler form and Euler density are related in the obvious way ${\mathcal E}_d = E_{d} \vol(M)$, for $\vol(M)$ the volume form on $M$.
In writing~\eqref{E:defEuler} we have normalized the Euler form so that its integral over an $S^d$ is $d!\Vol(S^d)$.

To define the Chern-Simons like boundary term, it is convenient to define a connection one-form and curvature two-form that interpolate linearly between a reference one-form $\omega_0$ and the actual one-form of interest $\omega$:
\begin{align}
\begin{split}
\omega(t)& \equiv t \omega+(1-t)\omega_0\,,
\\
\R(t)^A{}_B & \equiv \d\omega(t)^A{}_B + \omega(t)^A{}_C \wedge \omega(t)^C{}_B\,.
\end{split}
\end{align}
The boundary term is constructed from the $d-1$ form:
\be
\label{Qdef}
{\mathcal Q}_{d} \equiv \frac{d}{2}\int_0^1 \d t\, \dot \omega(t)^{A_1 A_2} \wedge \R(t)^{A_3 A_4} \wedge \cdots \wedge \R(t)^{A_{d-1} A_d} \epsilon_{A_1 \cdots A_d}\,.
\ee
(The density $Q_d$ is given by $\mathcal{Q}_d = Q_d \vol (\partial M)$.) If we also define 
\be
{\mathcal E}(t)_d \equiv \R(t)^{A_1 A_2} \wedge \cdots \wedge \R(t)^{A_{d-1} A_d}  \epsilon_{A_1 \cdots A_d} \ ,
\ee
then it follows, as we show below, 
\be
{\mathcal E}(1)_d - {\mathcal E}(0)_d =  \d {\mathcal Q}_{d} \ .
\label{EQrel}
\ee

The relevance of this construction to the Euler characteristic is that we can calculate the Euler characteristic for a manifold $M$ with boundary by comparing it to a manifold $M^0$ with the same boundary and zero Euler characteristic. Because $\chi(A\times B) = \chi(A)\chi(B)$ and because $\chi$ vanishes in odd dimensions, one such zero characteristic manifold is a product manifold where both $A$ and $B$ are odd dimensional. In a patch near the boundary, we can always choose to express the metric in Gaussian normal coordinates, 
\be
g= \d r^2 + f(r, x)_{\mu\nu} \d x^\mu \d x^\nu \,,
\ee
where the boundary is located at $r=r_0$. In this patch, we can choose a reference metric $g_0$ so that the patch is a product space, 
\be
g_0 = \d r^2 + f(r_0, x)_{\mu\nu} \d x^\mu \d x^\nu \ .
\ee
Let $\omega_0$ be the connection one-form associated with the metric $g_0$. By construction ${\mathcal E}_d(1) = {\mathcal E}_d$, and  it follows from the local relation (\ref{EQrel}) 
that the Euler characteristic for a manifold with boundary is 
\be
\label{Eulercharacteristic}
\chi(M) = \frac{2}{d! \Vol(S^d)} \left( \int_M {\mathcal E}_d- \int_{\partial M} {\mathcal Q}_d \right) \ .
\ee
We have normalized the characteristic so that
$\chi(S^d) = 2$.

On the boundary $\partial M$, we can give an explicit formula for $\dot{\omega}^{AB}$ in terms of the extrinsic curvature,
\be
\dot{\omega}(t)^{AB}=\omega^{AB}-\omega_0^{AB} = \mathcal{K}^A n^B - \mathcal{K}^B n^A\,,
\ee
where we have defined the extrinsic curvature one-form $\mathcal{K}_{\alpha} \equiv K_{\alpha\beta}\d y^{\beta}$, and converted its index to a flat index through the $e^A$, metric, and embedding functions. Similarly, $n^A = e^A_{\mu}n^{\mu}$.

In analogy with the two dimensional variation~\eqref{E:deltaW}, we therefore posit that the $a$-dependent piece of the Weyl 
anomaly is
\be
\delta_{\sigma} W = (-1)^{d/2} \frac{4a}{d! \Vol(S^d)}   \left( \int_M {\mathcal E}_d \delta \sigma - \int_{\partial M} {\mathcal Q}_d \delta \sigma \right) + \ldots 
\label{E:aAnomalyWithBdy}
\ee
where the ellipsis denotes terms depending on $c_i$, the total divergence in the trace anomaly, and possibly other purely boundary contributions. We verify this claim in subsection \ref{S:WZconsistency} by showing that the anomaly~\eqref{E:aAnomalyWithBdy} is Wess-Zumino consistent. With this variation in hand, we can integrate it in one of the same three ways we used in $d=2$: guess work, using the integral~\eqref{integratedanomaly}, or dimensional regularization. The integral~\eqref{integratedanomaly} gives the $a$ dependent contribution to the effective anomaly action,
\be
\label{E:integratedAnomalyGeneral}
\W[g_{\mu\nu},e^{-2\tau}g_{\mu\nu}] = (-1)^{d/2}\frac{4a}{d!\Vol(S^d)}\int_0^1 \d t \left.\left\{ \int_M\tau \mathcal{E}_d[g']- \int_{\partial M}\tau \mathcal{Q}_d[g']\right\}\right|_{g'_{\mu\nu}=e^{-2t\tau}g_{\mu\nu}}\,,
\ee  
We also deduce $\dW$ from dimensional regularization in subsection~\ref{S:dimReg}.

Let us next study the relation between ${\mathcal E}_d$ and ${\mathcal Q}_d$.
The relation~\eqref{EQrel} is an example of a ``transgression form'' (see e.g.~\cite{Jensen:2013kka} for a modern summary of transgression forms). To prove it, consider
\be
\label{E:dEdt}
\frac{\d}{\d t}\mathcal{E}(t)_d = \dot{\R}(t)^A{}_B\wedge \frac{\partial \mathcal{E}(t)_d}{\partial \R(t)^A{}_B}\,.
\ee
It is convenient to introduce an exterior covariant derivative ${\rm D}$.  
It takes tensor-valued $p$-forms to tensor-valued $p+1$-forms. For example it acts on a matrix-valued $p$-form, $f^A{}_B$ as
\be
{\rm D} f^A{}_B = \d f^A{}_B + \omega^A{}_C \wedge f^C{}_B - (-1)^p f^A{}_C \wedge \omega^C{}_B\,,
\ee
and correspondingly for (co)vector-valued forms. It has the Lifshitz property, e.g.
\be
\d(f^{AB}\wedge g_{AB}) ={\rm D}(f^{AB}\wedge g_{AB}) =  {\rm D} f^{AB} \wedge g_{AB} + (-1)^p f^{AB} \wedge {\rm D} g_{AB}\,.
\ee
Defining ${\rm D} (t)$, we then have
\be
{\rm D} (t)\R(t)^A{}_B = 0\,, \qquad \dot{\R}(t)^A{}_B  = {\rm D} (t) \dot{\omega}(t)^A{}_B\,.
\ee
The metric $\delta_{AB}$ and antisymmetric Levi-Civita tensor $\epsilon_{A_1\hdots A_d}$ are also constant under ${\rm D} (t)$, provided that we let the $e^A$ depend on $t$ so that $\omega(t)$ is associated with a metric $g(t)$. Consequently,
\be
{\rm D} (t) \frac{\partial \mathcal{E}(t)_d}{\partial \R(t)^{AB}} = \frac{d}{2}{\rm D}(t)\left(  \R(t)^{A_1A_2}\wedge \hdots \wedge \R(t)^{A_{d-3}A_{d-2}}\epsilon_{ABA_1\hdots A_d}\right)=0\,,
\ee
and we can rewrite~\eqref{E:dEdt} as
\begin{align}
\begin{split}
\frac{\d}{\d t}\mathcal{E}(t)_d &= \d \left(\dot{\omega}(t)^{AB}\wedge  \frac{\partial \mathcal{E}(t)_d}{\partial \R(t)^{AB}}\right)\,
\\
& = \d\left( \frac{d}{2}\dot{\omega}(t)^{A_1A_2}\wedge \R(t)^{A_3A_4}\wedge \hdots \wedge \R(t)^{A_{d-1}A_d}\varepsilon_{A_1\hdots A_d}\right)\,.
\end{split}
\end{align}
Integrating this equality over $t\in [0,1]$ immediately yields~\eqref{EQrel}.

\subsection{An Explicit Expression For The Boundary Term}
\label{S:explicitQd}

It will be expedient in the rest of this section to have an explicit expression for the boundary term $\int_{\partial M}\mathcal{Q}_d$, that is to perform the integral over $t$ in~\eqref{Qdef}. The final result is~\eqref{E:Qd}.

Before doing so, we will use that the pullback of $\R^{AB}$ to the boundary can be expressed in terms of the intrinsic and extrinsic curvatures of the boundary. 
The relations between $\R^{AB}$ and the boundary curvatures are known as the Gauss and Codazzi equations, and we discuss them in appendix \ref{app:diffgeom}.

Denoting the intrinsic Riemann curvature tensor of the boundary as $\oR^{\alpha}{}_{\beta\gamma\delta}$, we define the intrinsic curvature two-form
\be
\mathcal{\oR}^{\alpha}{}_{\beta} \equiv \frac{1}{2}\oR^{\alpha}{}_{\beta\gamma\delta}\d y^{\gamma}\wedge \d y^{\delta}\,,
\ee
and thereby $\mathcal{\oR}^A{}_B$. Using the boundary covariant derivative $\oD_{\alpha}$, we define a boundary exterior covariant derivative $\oD$ just like ${\rm D}$. The Gauss and Codazzi equations can then be summarized as
\be
\label{E:GaussAndCodazzi}
\R^A{}_B = \mathcal{\oR}^A{}_B - \mathcal{K}^A \wedge \mathcal{K}_B + n_B\oD \mathcal{K}^A  - n^A \oD \mathcal{K}_B\,,
\ee

We can similarly decompose the pullback of $\R(t)$. On the boundary
\be
\omega(t)^A{}_B = \omega^A{}_B + (t-1)(\mathcal{K}^A n_B- n^A\mathcal{K}_B) \ ,
\ee which implies that on the boundary
\begin{align}
\begin{split}
\R(t)^A{}_B &= \R^A{}_B + (t-1)\oD(\mathcal{K}^An_B -n^A\mathcal{K}_B) +(t-1)^2(\mathcal{K}^An_C-n^A\mathcal{K}_C)\wedge (\mathcal{K}^Cn_B - n^C \mathcal{K}_B)
\\
& = \R^A{}_B -(t^2-1)\mathcal{K}^A\wedge \mathcal{K}_B +(t-1)\left( n_B \oD\mathcal{K}^A - n^A\oD\mathcal{K}_B\right)\,,
\end{split}
\end{align}
where we have used that $\oD n^A = \mathcal{K}^A$. Putting this together with~\eqref{E:GaussAndCodazzi}, we have
\be
\R(t)^A{}_B = \mathcal{\oR}^A{}_B - t^2 \mathcal{K}^A\wedge \mathcal{K}_B +t\left( n_B \oD\mathcal{K}^A - n^A \oD\mathcal{K}_B\right)\,.
\ee

Then on the boundary the definition of $\mathcal{Q}_d$~\eqref{Qdef} becomes
\begin{align}
\nonumber
\mathcal{Q}_d&=d\int_0^1 \d t \,n^{A_1} \mathcal{K}^{A_2}\wedge \left( \mathcal{\oR}^{A_3A_4}-t^2\mathcal{K}^{A_3}\wedge\mathcal{K}^{A_4}\right)\wedge \hdots \wedge \left(\mathcal{\oR}^{A_{d-1}A_d}-t^2\mathcal{K}^{A_{d-1}}\wedge\mathcal{K}^{A_d}\right)\epsilon_{A_1\hdots A_d}
\\
& = d\sum_{k=0}^{m-1}\begin{pmatrix}m-1 \\ k \end{pmatrix}\frac{(-1)^k}{2k+1}\mathcal{\oR}^{m-1-k}\wedge \mathcal{K}^{2k+1}n^A \epsilon_{A\hdots}\,,
\label{E:Qd}
\end{align}
where we have defined $m\equiv \frac{d}{2}$ and in the last line we have suppressed the indices of the curvature forms, all of which are dotted into the epsilon tensor. We have also used that only one index of the epsilon tensor can be dotted into the normal vector $n^A$, and so the factors of $\oD\mathcal{K}^A$ in $\R(t)$ never appear in $\mathcal{Q}_d$.

The integral representation of $\mathcal{Q}_d$ in the first line of~\eqref{E:Qd} is not new. A similar expression appears in e.g. ref.\ \cite{Miskovic:2007mg}.

For example, in four and six dimensions we have
\be
\mathcal{Q}_4& =& 4 n^A\mathcal{K}^B\wedge \left( \mathcal{\oR}^{CD}-\frac{1}{3}\mathcal{K}^C\wedge \mathcal{K}^D\right)\epsilon_{ABCD}\,,
\\
\nonumber
\mathcal{Q}_6 & =& 6 n^A \mathcal{K}^B \wedge \left( \mathcal{\oR}^{CD}\wedge \mathcal{\oR}^{EF} - \frac{2}{3}\mathcal{\oR}^{CD}\wedge \mathcal{K}^E\wedge \mathcal{K}^F + \frac{1}{5}\mathcal{K}^B\wedge\mathcal{K}^C\wedge\mathcal{K}^D\wedge\mathcal{K}^E\wedge\mathcal{K}^F\right)\epsilon_{ABCDEF}\,.
\ee

\subsection{Wess-Zumino Consistency}
\label{S:WZconsistency}

We now verify that the posited term proportional to $a$ in the Weyl anomaly~\eqref{E:aAnomalyWithBdy} is Wess-Zumino consistent. In this setting, Wess-Zumino consistency requires that the anomaly satisfies
\be
[\delta_{\sigma_1},\delta_{\sigma_2}]W = 0\,.
\ee
Notating the anomalous variation proportional to $a$ as
\[
\delta_{\sigma}W_a = A \left( \int_M \delta \sigma \,\mathcal{E}_d - \int_{\partial M}\delta \sigma \, \mathcal{Q}_d \right) \,, \qquad A \equiv (-1)^{d/2}\frac{4a}{d!\Vol(S^d)}\,,
\]
we consider
\be
\delta_{\sigma_1} \delta_{\sigma_2}W_a = A \left( \int_M \delta \sigma_2  \delta_{\sigma_1}\mathcal{E}_d- \int_{\partial M} \delta \sigma_2  \delta_{\sigma_1}\mathcal{Q}_d\right)\,.
\ee

The variation of $\mathcal{E}_d$ is a total derivative,
\be
\label{E:deltaEuler}
\delta_{\sigma}\mathcal{E}_d = \d\left( \delta_{\sigma}\omega^{AB} \wedge \frac{\partial \mathcal{E}_d}{\partial \R^{AB}}\right)\,,
\ee
with
\be
\delta_{\sigma}\omega^{AB} =( e^A e^B_{\mu}-e^B e^A_{\mu})\partial^{\mu}\delta \sigma \,.
\ee
It then follows that the bulk part of the second variation is
\begin{align}
\nonumber
\delta_{\sigma_1}\delta_{\sigma_2}W_a &= 2dA \int_M e^{A_1}e^{A_2}_{\mu}\partial^{\mu}\delta \sigma_1  \wedge \d\delta \sigma_2 \wedge \mathcal{R}^{A_3A_4}\wedge \hdots \wedge \mathcal{R}^{A_{d-1}A_d}\epsilon_{A_1\hdots A_d} + (\text{boundary term})\,,
\\
&= A\int_M d^dx \sqrt{g} \, \mathcal{X}_d^{\mu\nu}\partial_{\mu}\delta \sigma_1 \partial_{\nu}\delta\sigma_2 + (\text{boundary term})\,,
\label{E:ddWBulk}
\end{align}
where we have defined
\be
\mathcal{X}_d^{\mu\nu} \equiv \frac{d}{2^{d/2}}R_{\nu_1\nu_2\rho_1\rho_2}\hdots R_{\nu_{d-3}\nu_{d-2}\rho_{d-3}\rho_{d-2}}\epsilon^{\mu\rho\nu_1\hdots \nu_{d-2}}\epsilon^{\nu}{}_{\rho}{}^{ \rho_1\hdots \rho_{d-2}}  \,.
\ee
$\mathcal{X}_d^{\mu\nu}$ is symmetric, $\mathcal{X}_d^{\mu\nu}=\mathcal{X}_d^{\nu\mu}$, on account of $R_{\mu\nu\rho\sigma}=R_{\rho\sigma\mu\nu}$. The symmetry of $\mathcal{X}_d^{\mu\nu}$ together with the variation \eqref{E:ddWBulk} imply
\be
[\delta_{\sigma_1},\delta_{\sigma_2}]W_a = (\text{boundary term})\,.
\ee
In other words, the bulk term in the $a$-anomaly is Wess-Zumino consistent. It suffices now to show that the boundary term also vanishes. 

To proceed, we require the Weyl variations of the extrinsic and intrinsic curvatures. The variation of $K_{\alpha\beta}$ and so $\mathcal{K}^A$ is
\be
\delta_{\sigma} K_{\alpha\beta} = \delta \sigma K_{\alpha\beta} + \gamma_{\alpha\beta} n^{\mu}\partial_{\mu}\delta \sigma \,, \qquad \delta_{\sigma}\mathcal{K}^A = e^An^{\mu}\partial_{\mu}\delta \sigma =( \delta_{\sigma}\omega^A{}_B) n^B\,,
\ee
where $e^A$ in the variation of $\mathcal{K}^A$ is pulled back to the boundary, while the variation of $\mathcal{\oR}^A{}_B$ is
\be
\delta_{\sigma}\mathcal{\oR}^A{}_B = \oD\delta_{\sigma}\mathring{\omega}^A{}_B\,,
\ee
for $\mathring{\omega}^A{}_B$ the connection one-form on the boundary. The variation of $\omega^A{}_B$ on the boundary is related to those of $\mathring{\omega}^A{}_B$ via
\be
\delta_{\sigma} \omega^A{}_B = \delta_{\sigma} \mathring{\omega}^A{}_B + \left( n_B\delta_{\sigma} \omega^A{}_C  - n^A \delta_{\sigma} \omega_{CB} \right)n^C\,.
\ee

Under a general variation of $\mathcal{K}^A$ and $\mathcal{\oR}^A{}_B$, $\mathcal{Q}_d$ in~\eqref{E:Qd} varies as
\be
\delta \mathcal{Q}_d = d \sum_{k=0}^{m-1}\begin{pmatrix} m-1 \\ k \end{pmatrix} (-1)^k\left\{  \delta \mathcal{K}^B \wedge\mathcal{\oR}^{CD}+\frac{m-1-k}{2k+1}\delta \mathcal{\oR}^{BC}\wedge  \mathcal{K}^D\right\} \wedge \mathcal{\oR}^{m-2-k}\wedge \mathcal{K}^{2k} n^A \epsilon_{ABCD\hdots}\,.
\ee
Specializing to Weyl variations, this becomes
\begin{align}
\begin{split}
\frac{1}{d}\delta_{\sigma}\mathcal{Q}_d = & \delta_{\sigma}\omega^B{}_C n^C \R^{m-1} n^A \epsilon_{AB\hdots} 
\\
& \qquad + \delta_{\sigma} \mathring{\omega}^{BC} \wedge \sum_{k=0}^{m-2} \begin{pmatrix}m-2 \\ k \end{pmatrix} (-1)^k(m-1) \oD\mathcal{K}^D \wedge \mathcal{\oR}^{m-2-k}\wedge \mathcal{K}^{2k} n^A \epsilon_{ABCD\hdots}
\\
& \qquad + \d\left\{ \delta_{\sigma}\mathring{\omega}^{BC} \wedge \sum_{k=0}^{m-2} \begin{pmatrix} m-2 \\ k \end{pmatrix} (-1)^k\frac{m-1}{2k+1} \mathcal{\oR}^{m-2-k}\wedge \mathcal{K}^{2k+1}n^A \epsilon_{ABC\hdots}\right\}\,,
\end{split}
\end{align}
where we have used the Gauss equation in simplifying the $\delta_{\sigma} \mathcal{K}$ variation along with $\oD \mathcal{\oR}=0$ in simplifying the $\delta_{\sigma}\mathcal{\oR}$ variation. Using the Codazzi equation, $\R^A{}_B n^B = \oD\mathcal{K}^A$, the second line combines with the first to give
\be
\label{E:deltaWeylQd}
\delta_{\sigma}\mathcal{Q}_d =  \delta_{\sigma}\omega^{AB} \wedge \frac{\partial \mathcal{E}_d}{\partial \R^{AB}}+ \d \left\{ (m-1)\delta_{\sigma}\mathring{\omega}^{AB} \wedge( \mathcal{Q}_{d-2})_{AB}\right\}\,.
\ee
In writing the boundary term, we have defined the matrix-valued $(d-3)$-form $(\mathcal{Q}_{d-2})_{AB}$ to be
\be
\label{E:QdM2}
(\mathcal{Q}_{d-2})_{AB} \equiv d\sum_{k=0}^{m-2} \begin{pmatrix} m-2 \\ k \end{pmatrix} \frac{(-1)^k}{2k+1} \mathcal{\oR}^{m-2-k}\wedge \mathcal{K}^{2k+1}n^C \epsilon_{ABC\hdots}\,
\ee
The reason for the name is the similarity with the explicit expression~\eqref{E:Qd} for $\mathcal{Q}_d$: the sum~\eqref{E:QdM2} is identical to that in the expression for $\mathcal{Q}_d$, except it runs to $k=m-2$ rather than $k=m-1$.

Putting $\delta_\sigma \mathcal{Q}_d$ together with the variation of the Euler form~\eqref{E:deltaEuler}, the boundary term in the variation of $\int_M \delta \sigma_2 \delta_{\sigma_1} \mathcal{E}_d$ cancels against the first half of the variation of $\mathcal{Q}_d$ in~\eqref{E:deltaWeylQd}, so that
\begin{align}
\begin{split}
\label{E:totalddWa}
\delta_{\sigma_1}\delta_{\sigma_2}W_a = & A\left(  \int_M d^dx \sqrt{g}\, \mathcal{X}^{\mu\nu}\partial_{\mu}\delta\sigma_1 \partial_{\nu}\delta \sigma_2 -2(m-1)\int_{\partial M} e^A e^B_{\alpha}\partial^{\alpha}\delta \sigma_1 \wedge \d \delta \sigma_2 \wedge (\mathcal{Q}_{d-2})_{AB} \right)
\\
= & A \left( \int_M d^dx \sqrt{g}\, \mathcal{X}^{\mu\nu}\partial_{\mu}\delta\sigma_1 \partial_{\nu}\delta \sigma_2 - \int_{\partial M}d^{d-1}y\sqrt{\gamma}\, \mathcal{Y}^{\alpha\beta}\partial_{\alpha}\delta \sigma_1 \partial_{\beta}\delta\sigma_2\right)\,,
\end{split}
\end{align}
where $\mathcal{Y}^{\alpha \beta}$ is 
\begin{align}
\begin{split}
\mathcal{Y}^{\alpha\beta} = &d\epsilon^{\alpha\gamma\gamma_1\hdots \gamma_{d-3}}\epsilon^{\beta}{}_{\gamma}{}^{\delta_1\hdots \delta_{d-3}} \sum_{k=0}^{m-2}\begin{pmatrix} m-2\\ k\end{pmatrix} (-1)^k\frac{m-1}{(2k+1)2^{m-3-k}}
\\
& \qquad\times  \oR_{\gamma_1\gamma_2\delta_1\delta_2} \cdots \oR_{\gamma_{d-2k-5}\gamma_{d-2k-4}\delta_{d-2k-5}\delta_{d-2k-4}} K_{\gamma_{d-2k-3}\delta_{d-2k-3}}\cdots K_{\gamma_{d-3}\delta_{d-3}}\,.
\end{split}
\end{align}
$\mathcal{Y}^{\alpha\beta}$ is symmetric owing to the symmetry of the boundary curvatures, $\oR_{\alpha\beta\gamma\delta}=\oR_{\gamma\delta\alpha\beta}$ and $K_{\alpha\beta}=K_{\beta\alpha}$. Then~\eqref{E:totalddWa} yields
\be
[\delta_{\sigma_1},\delta_{\sigma_2}]W_a=0\,,
\ee
which is what we sought to show.

\subsection{A Complete Classification in $d=4$ and Boundary Central Charges}
\label{S:classify4d}

The previous subsection was somewhat abstract. Let us see how the consistency works in $d=4$. Along the way, we will also classify the potential boundary terms in the Weyl anomaly, finding two ``boundary central charges.'' To our knowledge, one of these ``central charges'' was first noted in~\cite{Melmed} and the other later in ref.\ \cite{Dowker:1989ue}.

In $d=4$, $\mathcal{E}_4$ and $\mathcal{Q}_4$ are equivalent to the scalars
\begin{align}
\begin{split}
E_4 &= R_{\mu\nu\rho\sigma}R^{\mu\nu\rho\sigma} - 4 R_{\mu\nu}R^{\mu\nu}+R^2\,,
\\
Q_4 & = 4\left(2 \mathring{E}_{\alpha\beta}K^{\alpha\beta} +\frac{2}{3}\tr (K^3) - K K_{\alpha\beta}K^{\alpha\beta} + \frac{1}{3}K^3\right)\,,
\end{split}
\end{align}
where $\mathring{E}_{\alpha\beta}=\oR_{\alpha\beta} - \frac{\oR}{2}\gamma_{\alpha\beta}$ is the boundary Einstein tensor, and the $a$-type term in the anomaly is
\be
\delta_{\sigma}W_a = A \left( \int_M \d^4x \sqrt{g} \, \delta \sigma E_4 - \int_{\partial M} \d^3y \sqrt{\gamma}\, \delta \sigma Q_4\right)\,, \qquad A = \frac{a}{16\pi^2}\,.
\ee
The Weyl variations of $E_4$ and $Q_4$ are
\begin{align}
\delta_{\sigma}E_4 &= - 4\delta \sigma E_4 + 8 {\rm D}_{\mu}\left( E^{\mu\nu}\partial_{\nu}\delta \sigma\right)\,,
\\
\nonumber
\delta_{\sigma}Q_4 & = - 3 \delta \sigma Q_4-4\left\{ R^{\alpha\beta}{}_{\alpha\beta} n^{\mu}\partial_{\mu} - 2\oD_{\alpha}\left(  K^{\alpha\beta}-K\gamma^{\alpha\beta}\right)\oD_{\beta} \right\} \delta \sigma - 8 \oD_{\alpha}\left\{ \left( K^{\alpha\beta}-K\gamma^{\alpha\beta}\right) \partial_{\beta}\delta\sigma\right\}\,,
\end{align}
Using the Gauss and Codazzi equations~\eqref{E:GaussAndCodazzi}, which here are
\be
\label{E:GaussCodazzi2}
R_{\alpha\beta\gamma\delta} = \oR_{\alpha\beta\gamma\delta} - K_{\alpha\gamma}K_{\beta\delta}+K_{\alpha\delta}K_{\beta\gamma}\,, \qquad n^{\mu}R_{\mu\alpha\beta\gamma} =  \oD_{\gamma}K_{\alpha\beta} - \oD_{\beta}K_{\alpha\gamma}\,,
\ee
we can rewrite the variation of $Q_4$ as
\be
\delta_{\sigma}Q_4 = -3 \delta \sigma Q_4 + 8 n_{\mu}E^{\mu\nu}\partial_{\nu}\delta \sigma -8 \oD_{\alpha}\left\{ \left( K^{\alpha\beta}-K\gamma^{\alpha\beta}\right)\partial_{\beta}\delta\sigma\right\}\,.
\ee
The second variation of $W_a$ is then
\be
\delta_{\sigma_1}\delta_{\sigma_2}W_a = -8A \left( \int_M \d^4x \sqrt{g}\, E^{\mu\nu}(\partial_{\mu}\delta \sigma_1)( \partial_{\nu}\delta\sigma_2)+ \int_{\partial M}\d^3y \sqrt{\gamma}\left( K^{\alpha\beta}-K\gamma^{\alpha\beta}\right)( \partial_{\alpha}\delta\sigma_1)( \partial_{\beta}\delta \sigma_2)\right)\,,
\ee
which is manifestly symmetric under $\delta\sigma_1\leftrightarrow\delta\sigma_2$, so that
\be
[\delta_{\sigma_1},\delta_{\sigma_2}]W_a = 0\,.
\ee
In this instance, the tensors $\mathcal{X}^{\mu\nu}$ and $\mathcal{Y}^{\alpha\beta}$ are
\be
\mathcal{X}^{\mu\nu} = - 8 E^{\mu\nu}\,, \qquad \mathcal{Y}^{\alpha \beta} = 8\left( K^{\alpha\beta}-K\gamma^{\alpha\beta}\right)\,.
\ee

So much for showing that the $a$-type anomaly is consistent. Are there any other boundary terms which may be allowed in the anomaly? This is essentially a cohomological question, which we answer in three steps:
\begin{enumerate}
\item Posit the most general boundary variation of $W$ characterized by dimensionless coefficients.
\item Use the freedom to add local boundary counterterms to remove as many of these coefficients as possible.
\item Demand that the residual variation is Wess-Zumino consistent.
\end{enumerate}

We perform this algorithm in Appendix~\ref{A:WZ4d}. The final result is that the total Weyl anomaly for a $d=4$ CFT is
\be
\label{E:4dTotalAnomaly}
\delta_{\sigma}W = \frac{1}{16 \pi^2} \int_M \d^4x\sqrt{g} \,\delta\sigma \left( a E_4 - c W_{\mu\nu\rho\sigma}^2\right) - \int_{\partial M}\d^3y\sqrt{\gamma}\,\delta\sigma\left( A Q_4 - b_1 \text{tr}\hat{K}^3-b_2 \gamma^{\alpha\gamma}\hat{K}^{\beta\delta}W_{\alpha\beta\gamma\delta}\right)\,,
\ee
where $\hat{K}_{\alpha\beta}$ is the traceless part of the extrinsic curvature, $\hat{K}_{\alpha\beta}=K_{\alpha\beta} - \frac{K}{d-1}\gamma_{\alpha\beta}$, and $W_{\alpha\beta\gamma\delta}$ is the pullback of the Weyl tensor. The terms proportional to $b_1$ and $b_2$ are the additional type-B boundary terms in the anomaly. We refer to $b_1$ and $b_2$ as ``boundary central charges,'' and they are formally analogous to $c$ 
insofar as they multiply Weyl-covariant scalars. The purely extrinsic term proportional to $b_1$ first appeared in~\cite{Melmed}, and the second term proportional to $b_2$ later appeared in \cite{Dowker:1989ue}.

It is an interesting exercise to compute $b_1$ and $b_2$ for a conformally coupled scalar field.  The simplest way to proceed is to look at existing heat kernel calculations for a scalar field in the presence of a boundary and then restrict to the conformally coupled case.  The action for such a conformally coupled scalar is
\be
S = \int_M \d^4 x \sqrt{g}\left( (\partial \phi)^2 + \frac{1}{6} R \phi^2 \right) + \frac{1}{3} \int_{\partial M} \d^3 y\sqrt{\gamma} K \phi^2 \ .
\ee
Note that the last term ensures Weyl invariance.  It is also necessary for coupling the theory to gravity.\footnote{%
If we are not interested in dynamical gravity, we could add an additional boundary term of the form $\phi (K + 3 n^\mu \partial_\mu )  \phi$ with arbitrary coefficient. This term preserves Weyl invariance.  However, it does not modify the boundary conditions or the scalar functional determinant. Consequently the 
boundary central charges that we determine below do not depend on this term. See the appendix of~\cite{Jensen:2015swa} for a related discussion.
}
By comparing this result with heat kernel calculations for a conformally coupled scalar field in the presence of a boundary, we can extract values for $b_1$ and $b_2$. There are two Weyl-invariant boundary conditions to consider, Dirichlet $\phi|_{\partial M} = 0$ (in which case the boundary term can be neglected) and the conformally-invariant Robin $(n^\mu \partial_\mu  + \frac{1}{3} K) \phi |_{\partial M} = 0$.  Comparing with for example (1.17) of \cite{McAvity:1992fq} or the expressions for $a_4$ on p 5 of \cite{Branson:1995cm}, we deduce that
\be
b_1({\rm Robin}) = - \frac{1}{(4 \pi)^2} \frac{2}{45} \ , \; \; \;
b_1({\rm Dirichlet}) = - \frac{1}{(4 \pi)^2} \frac{2}{35} \ , \; \; \;
b_2 = \frac{1}{(4\pi)^2} \frac{1}{15} \ .
\ee
The value for $b_1({\rm Dirichlet})$ was computed before in eq.\ (19) of ref.\ \cite{Melmed}, while $b_1({\rm Robin})$ can be found in eq.\ (55) of ref.\ \cite{IGMoss}. The coefficient $b_2$ was computed in the Dirichlet case in eq.\ (15) of ref.\ \cite{Dowker:1989ue}.  (In our conventions, $a = 1/360$ and $c= 1/120$ for a $4d$ conformally coupled scalar.) As $|b_1({\rm Dirichlet})| > |b_1({\rm Robin})|$, and one can flow from the Robin theory to the Dirichlet theory by deforming the Robin theory by a ``boundary mass'' $\int \d^3y\, m \phi^2$; it is tempting to speculate that $b_1$ satisfies a monotonicity property under boundary renormalization group flows, 
similar to the one conjectured for $a$ by Cardy and now proven in $d=4$ by ref.\ \cite{Komargodski:2011vj}. This conjecture is different from the ``boundary $F$-theorem'' conjectured in~\cite{Jensen:2013lxa,Estes:2014hka,Gaiotto:2014gha} for $d=4$ boundary flows. We leave further analysis of these boundary central charges $b_1$ and $b_2$ for the future.

\subsection{Dimensional Regularization}
\label{S:dimReg}

In the two dimensional case, we saw that an effective anomaly action could be constructed in dimensional regularization using a combination of the Einstein-Hilbert action and the Gibbons-Hawking surface term in $n = 2 + \epsilon$ dimensions. In the limit $\epsilon \to 0$, these objects sum together to give the Euler characteristic.  The obvious guess, which we shall verify, is that to construct the anomaly action in $d$ dimensions, we need to continue the Euler density along with the ${\mathcal Q}_d$ Chern-Simons like term to $n = d+\epsilon$ dimensions.  In the mathematics community, such a dimensionally continued Euler density is called a Lipschiftz-Killing curvature, while in the physics community, these objects are used to construct actions for Lovelock gravities.  

The $m$th Lipschitz-Killing curvature form in dimension $n$, $2m \leq n$, is:
\be
{\mathcal E}_{n,m} \equiv \left( \bigwedge_{i=1}^{m} \R^{A_{2i-1}A_{2i}} \right) \wedge \left( \bigwedge_{i=2m+1}^n e^{A_i} \right) \epsilon_{A_1 \cdots A_n} \ ,
\label{E:nm}
\ee
where $\epsilon_{A_1 \cdots A_n}$ is the totally antisymmetric Levi-Civita tensor in dimension $n$.  In $n=2m$ dimensions, the Lipschitz-Killing form reduces to the Euler form, ${\mathcal E}_{2m,m} = {\mathcal E}_{2m}$.  The analog of the Gibbons-Hawking term we call ${\mathcal Q}_{n,m}$:
\be
\label{Qdef2}
{\mathcal Q}_{n,m} \equiv  m \int_0^1  \, \dot \omega(t)^{A_1 A_2} \wedge \left( \bigwedge_{i=2}^{m} \R(t)^{A_{2i-1} A_{2i}} \right) \wedge
\left( \bigwedge_{i=2m+1}^n e^{A_i} \right) \epsilon_{A_1 \cdots A_n} \, \d t \ .
\ee
It is a $n-1$ degree Chern-Simons like form which is only defined on the boundary, which reduces to ${\mathcal Q}_d$ in $n=2m$ dimensions.

The obvious guess for the effective action $\widetilde W[g_{\mu\nu}]$ in $n = d+\epsilon$ dimensions, i.e.\
the $d$ dimensional analog of~\eqref{tildeWtwo}, is 
\be
\label{tildeWgeneral}
\widetilde W[g_{\mu\nu}] =
(-1)^{m}\frac{ 4 a}{(n-2m) (2m)! \Vol(S^{2m})} \left(  \int_M {\mathcal E}_{n,m} - \int_{\partial M} {\mathcal Q}_{n, m} \right) \ , 
\ee
where $d= 2m$. The effective anomaly action is then just
\be
\dW[g_{\mu\nu}, e^{-2 \tau} g_{\mu\nu}] = \lim_{n\to d} \left(\widetilde W[g_{\mu\nu}] - \widetilde W[e^{-2 \tau} g_{\mu\nu}]\right) \ .
\label{E:dWgeneral}
\ee
Note that this effective action only recovers the $a$ dependent portion of the trace anomaly.  

As in subsection~\ref{S:explicitQd}, we can perform the integral over $t$ in the definition of $\mathcal{Q}_{n,m}$ to deduce an explicit expression for $\mathcal{Q}_{n,m}$ in terms of the extrinsic and intrinsic curvatures of the boundary. The integration over $t$ is identical to that performed in subsection~\ref{S:explicitQd}, except now we have $n-2m$ factors of $e^A$ to account for. The final result is
\be
\mathcal{Q}_{n,m} = 2m \sum_{k=0}^{m-1}\begin{pmatrix} m-1 \\ k \end{pmatrix} \frac{(-1)^k}{2k+1}\mathcal{\oR}^{m-1-k}\wedge\mathcal{K}^{2k+1}\wedge e^{n-2m} n^A \epsilon_{A\hdots}\,,
\label{E:Qnm}
\ee
where for brevity we have suppressed the indices of the curvatures and factors of $e^A$, all of which are contracted with the remaining indices of the epsilon tensor.

Next we show that dimensional regularization~\eqref{tildeWgeneral} reproduces the $a$ portion of the Weyl anomaly. Our approach is almost identical to the demonstration that the $a$-anomaly is Wess-Zumino consistent in subsection~\ref{S:WZconsistency}. We begin with the expressions~\eqref{E:nm} and~\eqref{E:Qnm} for $\mathcal{E}_{n,m}$ and ${\mathcal Q}_{n,m}$. We consider the Weyl variation of
\be
\int_M \mathcal{E}_{n,m} - \int_{\partial M}\mathcal{Q}_{n,m} \, ,
\ee
in $n$ dimensions. We compute this variation in two steps. First we show that this difference does not depend on any variation of the connection one-form $\omega^A{}_B$ while keeping the $e^A$ fixed.\footnote{This same computation shows that the Lovelock gravities have a well-defined variational principle for the metric $g_{\mu\nu}$ on a space with boundary (see ref.\ \cite{Myers:1987yn}).} Then the Weyl variation only arises from the Weyl variation of the $e^A$ while keeping the $\omega^A{}_B$ fixed. This last variation is rather simple, as the $e^A$ only appear through wedge products in $\mathcal{E}_{n,m}$ and $\mathcal{Q}_{n,m}$.

Now consider a variation of the connection one-form $\omega^A{}_B$ whilst keeping the $e^A$ and embedding of the boundary fixed. The bulk and boundary curvatures vary as
\be
\delta_{\omega} \R^A{}_B = {\rm D} \delta \omega^A{}_B\,, \qquad \delta_{\omega} \mathcal{\oR}^A{}_B = \oD \delta \mathring{\omega}^A{}_B\,, \qquad \delta_{\omega} \mathcal{K}^A = (\delta \omega^A{}_B) n^B\,,
\ee
where $\mathring{\omega}^A{}_B$ is the connection one-form on the boundary. The computation of this variation is virtually identical to that in subsection~\ref{S:WZconsistency}, as the only difference between $\mathcal{E}_{n,m}$ and $\mathcal{E}_d$, and $\mathcal{Q}_{n,m}$ and $\mathcal{Q}_d$, is an extra wedge product of $n-2m$ factors of the $e^A$. The analogues of~\eqref{E:deltaEuler} and~\eqref{E:deltaWeylQd} are
\begin{align}
\begin{split}
\delta_{\omega} \mathcal{E}_{n,m} &= \d \left( \delta \omega^{AB} \wedge \frac{\partial \mathcal{E}_{n,m}}{\partial\R^{AB}}\right)\,,
\\
\delta_{\omega} \mathcal{Q}_{n,m} & = \delta \omega^{AB} \wedge \frac{\partial \mathcal{E}_{n,m}}{\partial\R^{AB}} + (\text{total deriative})\,,
\end{split}
\end{align}
so that
\be
\delta_{\omega}\left( \mathcal{E}_{n,m}-\d\mathcal{Q}_{n,m}\right)=0\,,
\ee
as claimed.

Now consider a variation under which $\omega^A{}_B$ is fixed and the $e^A$ vary as in an infinitesimal Weyl rescaling,
\be
\delta_{\sigma}e^A = \delta \sigma e^A\,.
\ee
Then
\be
\delta_{\sigma}\left( \mathcal{E}_{n,m} - \d \mathcal{Q}_{n,m}\right) = (n-2m)\delta\sigma \left( \mathcal{E}_{n,m} - \d \mathcal{Q}_{n,m}\right)\,, 
\ee
so that the variation of the dimensionally regulated anomaly action $\widetilde{W}$ in~\eqref{tildeWgeneral} is
\be
\delta_{\sigma}\widetilde{W} = (-1)^m \frac{4a}{(2m)!\Vol(S^{2m})}\left( \int_M \mathcal{E}_{n,m}\delta \sigma - \int_{\partial M}\mathcal{Q}_{n,m}\delta \sigma\right)\,.
\ee
In the $n\to 2m$ limit, this variation coincides with the $a$-anomaly~\eqref{E:aAnomalyWithBdy}.

\section{Dilaton Effective Actions and Boundary Terms}
\label{sec:dilaton}

In this section, we present the $a$ contribution  to the dilaton effective action in a spacetime with boundary in four and six dimensions. The $d=2$ dilaton effective action with a bounday term is given by \eqref{Wefftwod}. For $d>2$, the computation of boundary terms is more laborious. The details of a derivation using dimensional regularization are provided in appendix \ref{app:dimreg} in dimensions four and six. We save the general discussion of how the universal entanglement entropy arises from the boundary terms of these dilaton actions for the next section.

\subsection{The Dilaton Effective Action in $d=4$}

\label{S:dilaton4d}

The Euler density in $d=4$ is given by 
\be
E_4={1\over 4} \delta^{\mu_1 \cdots \mu_4}_{\nu_1 \cdots \nu_4} R^{\nu_1 \nu_2}{}_{\mu_1 \mu_2} R^{\nu_3 \nu_4}{}_{\mu_3 \mu_4} = R_{\mu\nu\rho\sigma}R^{\mu\nu\rho\sigma} - 4 R_{\mu\nu}R^{\mu\nu} + R^2\,,
\ee
where $\delta^{\mu_1\cdots \mu_4}_{\nu_1\cdots \nu_4}$ is the fully antisymmetrized product of four Kronecker delta functions. The boundary term is
\be
Q_4=-4 \delta^{\mu_1\mu_2\mu_3}_{\nu_1\nu_2\nu_3}~ K^{\nu_1}_{\mu_1}\left({1\over 2} R^{\nu_2 \nu_3}{}_{\mu_2 \mu_3} + {2\over 3} K^{\nu_2}_{\mu_2} K^{\nu_3}_{\mu_3} \right) =4\left( 2 \mathring{E}_{\alpha\beta}K^{\alpha\beta} + \frac{2}{3}\text{tr}(K^3) - K K_{\alpha\beta}K^{\alpha\beta}+\frac{1}{3}K^3\right)\ .
\ee
Denote the Einstein tensor as 
\be
E^{\mu\nu}=R^{\mu\nu}- {1\over 2}g^{\mu\nu}R\,.
\ee 
In appendix \ref{app:dimreg}, we find the dilaton effective action in $d=4$ to be
\begin{eqnarray}
\W[g_{\mu\nu}, e^{-2 \tau} g_{\mu\nu}] &=&
 \frac{a}{(4\pi)^2} \int_M \d^4 x \sqrt{g} \left[ \tau E_4 + 4 E^{{\mu\nu}} (\partial_\mu \tau) (\partial_\nu \tau)
+ 8 ({\rm D}_\mu \partial_\nu \tau)(\partial^\mu \tau)( \partial^\nu \tau) + 2 (\partial \tau)^4 \right] \nonumber \\
&&
-\frac{a}{(4\pi)^2} \ \int_{\partial M} \d^3 y\sqrt{\gamma} \left[ \tau Q_4 + 4 (K \gamma^{\alpha\beta} - K^{\alpha\beta}) (\partial_\alpha \tau)(\partial_\beta \tau) + \frac{8}{3} \tau_n^3 \right]  \ ,
\label{Weffdfour}
\end{eqnarray}  
where $\tau_n = n^\mu \partial_\mu \tau$ is a normal derivative of the Weyl scale factor.  
The bulk term agrees with ref.\ \cite{Komargodski:2011vj, BrownOttewill} while the boundary contribution is to our knowledge a new result.

\subsection{The Dilaton Effective Action in $d=6$}

The Euler density in $d=6$ is given by
\be
\label{e6}
E_6={1\over 8}  \delta^{\mu_1 \cdots \mu_6}_{\nu_1 \cdots \nu_6} R^{\nu_1 \nu_2}{}_{\mu_1 \mu_2} R^{\nu_3 \nu_4}{}_{\mu_3 \mu_4} R^{\nu_5 \nu_6}{}_{\mu_5 \mu_6} 
\ee
and the boundary term is
\begin{align}
\begin{split}
\label{q6}
Q_6=&-6\delta^{\beta_1 \cdots \beta_5}_{\alpha_1\cdots \alpha_5}  K^{\alpha_1}_{\beta_1} \left[\left({1\over 2} R^{\alpha_2 \alpha_3}{}_{\beta_2 \beta_3}+ 
 {2\over 3} K^{\alpha_2}_{\beta_2} K^{\alpha_3}_{\beta_3}\right) \left({1\over 2} R^{\alpha_4 \alpha_5}{}_{\beta_4 \beta_5} +  {2\over 3} K^{\alpha_4}_{\beta_5} K^{\alpha_4}_{\beta_5}\right)\right.
 \\
 & \qquad \qquad \qquad \qquad \left. + 
     {4\over 45} K^{\alpha_2}_{\beta_2} K^{\alpha_3}_{\beta_3}   K^{\alpha_4}_{\beta_5} K^{\alpha_4}_{\beta_5}\right]  \ .
\end{split}
\end{align}

To present the bulk dilaton action, we define
\begin{align}
\begin{split}
\label{E2def}
E^{(2)\mu\nu}&\equiv g^{\mu \nu} E_4+ 8  R^\mu_\rho  R^{\rho \nu } - 4  R^{\mu \nu}  R + 8  R_{\rho\sigma}   R^{\mu\rho\nu\sigma} - 4 R^{\mu}{}_{\rho \sigma \tau}  R^{\nu \rho \sigma \tau} \ ,  
\\
C_{\mu\nu\rho\sigma}&\equiv R_{\mu\nu\rho\sigma} -g_{\mu \rho} R_{\nu \sigma} + g_{\mu\sigma} R_{\nu \rho}  \ . 
\end{split}
\end{align}
In appendix \ref{app:dimreg}, we use dimensional regularization to find the bulk dilaton action 
\begin{align}
\begin{split}
\label{Weffdsixbulk}
\W[g_{\mu\nu}, e^{-2 \tau} g_{\mu\nu}]_{(\rm{Bulk})}=&
\\
 {a \over 3 (4\pi)^3} \int_M \d^6 x \sqrt{g}&\left\{- \tau E_6 + 3 E^{(2)}_{\mu\nu} \partial^\mu \tau \partial^\nu \tau+ 16 C_{\mu\nu\rho\sigma} ({\rm D}^\mu \partial^\rho \tau) (\partial^\nu \tau)( \partial^\sigma \tau) \right.
\\
& + 16 E^{}_{\mu\nu} \left[ (\partial^\mu \tau) (\partial^\rho \tau) ({\rm D}_\rho\partial^\nu \tau) - (\partial^\mu \tau) (\partial^\nu \tau) \Box \tau \right]- 6 R (\partial \tau)^4 
\\
&\left. - 24 (\partial \tau)^2 ({\rm D} \partial \tau)^2 + 24 (\partial \tau)^2 (\Box \tau)^2 - 36 (\Box \tau) (\partial \tau)^4 + 24 (\partial \tau)^6\right\} \ .
\end{split}
\end{align}
This reproduces the bulk Wess-Zumino term first obtained in \cite{Elvang:2012st}. 

We have not been able to generate the boundary term in a general curved background.  However, for a conformally flat geometry, we find 
\begin{align}
\label{6dresult}
\W[\delta_{\mu\nu}, e^{-2 \tau} \delta_{\mu\nu}]=-{a \over 16\pi^3}& \int_M \d^6 x \sqrt{g}\left\{2 (\partial \tau)^2 (\partial_\mu \partial_\nu \tau)^2 -2 (\partial \tau)^2 (\Box \tau)^2 +3 \Box \tau (\partial \tau)^4 -2(\partial \tau)^6\right\} \nn
\\
-{a \over 3 (4\pi)^3} \int_{\partial M} \d^5 y\sqrt{\gamma}&\Big[
-\tau  Q_6[\delta_{\mu\nu}]
+48 P^\alpha_\beta (\partial_\alpha \tau) (\partial^\beta \tau) 
+ 3 Q_4[\delta_{\mu\nu}] (\oD \tau)^2 \nn 
\\
&
+ 48   K^{\alpha \beta}   (\oBox \tau)  (\oD_\alpha \partial_\beta \tau)  
+ 24 K (\oD_\alpha \partial_\beta \tau)^2 
- 48  K_{\alpha \gamma} (\oD^\beta \partial^\alpha \tau) (\oD^\gamma \partial_\beta \tau) \nn
\\
&
-24 K (\oBox \tau)^2 
-32 K (\oD \tau)^2 \oBox \tau
-16 K (\partial^\alpha \tau) (\partial^\beta \tau) (\oD_\alpha \partial_\beta \tau)
\\
&
+16 K_{\alpha \beta} (\partial^\alpha \tau) (\partial^\beta \tau)\oBox \tau
+32  K_{\alpha \beta} (\oD^\alpha \partial^\beta \tau) (\oD \tau)^2
+ 12 K \tau_n^4   \nn
\\
&
+ 12 K (\oD\tau)^4 
+ 24 K  (\oD\tau)^2  \tau_n^2 
+ 48 (\oBox \tau) (\oD\tau)^2  (\tau_n)
+ 16 (\oBox \tau) (\tau_n^3)  \nn
\\
&
- 24  (\oD\tau)^2 \tau_n^3
- 36 \tau_n (\oD\tau)^4
- {36\over 5} \tau_n^5\Big] \nn\,,
\end{align}
where we have defined
\begin{eqnarray}
\label{Pabdef}
P^\alpha_\beta &\equiv& \left(K^2   - \text{tr}(K^2)\right)K^\alpha_\beta  -2 K  K^{\alpha \gamma}  K_{\beta \gamma} + 2  K_{\gamma \delta} K^{\alpha \gamma} K^{\delta}_\beta \ .
\end{eqnarray}

\section{The Sphere Entanglement Entropy: General Result}
\label{sec:example}

We consider the entanglement entropy across a sphere with radius $\ell$ in flat space.  The calculation is analogous to the discussion of the entanglement entropy for an interval in $d=2$ in section \ref{sec:planetocylinder}.  The information necessary to compute the entropy is contained in the causal development of the interior of the sphere, a ball of radius $\ell$.  We can then map that causal development to all of hyperbolic space cross the real line $R \times H^{d-1} $ using the change of variables
\begin{align}
\begin{split}
t &= \ell \frac{\sinh \tau/\ell}{\cosh u + \cosh \tau/\ell} \ , \\
r &= \ell \frac{\sinh u}{\cosh u + \cosh \tau/\ell} \ , 
\end{split}
\end{align}
where $\tau$ labels the new time, $u$ is the radial coordinate in hyperbolic space while $(t,r)$ are time and radius in polar coordinates in flat space. The line elements on flat space and $R\times H^{d-1} $ are related by a Weyl rescaling 
(see for example ref.\ \cite{Candelas:1978gf})
\begin{align}
\begin{split}
\eta  &= -\d t^2 + \d r^2 + r^2 \d \Omega_{d-2}^2 \ , \\
&= e^{2 \sigma} \left[ -\d \tau^2 + \ell^2 (\d u^2 + \sinh^2 u \, \d \Omega_{d-2}^2) \right] \ ,
\label{HSmetric}
\end{split}
\end{align}
where
$
 e^{-\sigma} = \cosh u + \cosh \tau/\ell 
$.
We proceed by using the Euclidean version of this map, where $\tau_E$ is a periodic variable with period $2\pi \ell$ so that the theory is naturally at a temperature $T = \frac{1}{2\pi \ell}$, and the Euclidean geometry is conformal to $S^1\times\mathbb{H}^{d-1}$. Note a difference here with the $d=2$ case where the temperature was a free parameter. 

\begin{figure}
\begin{center}
a) \includegraphics[width = 2.5in]{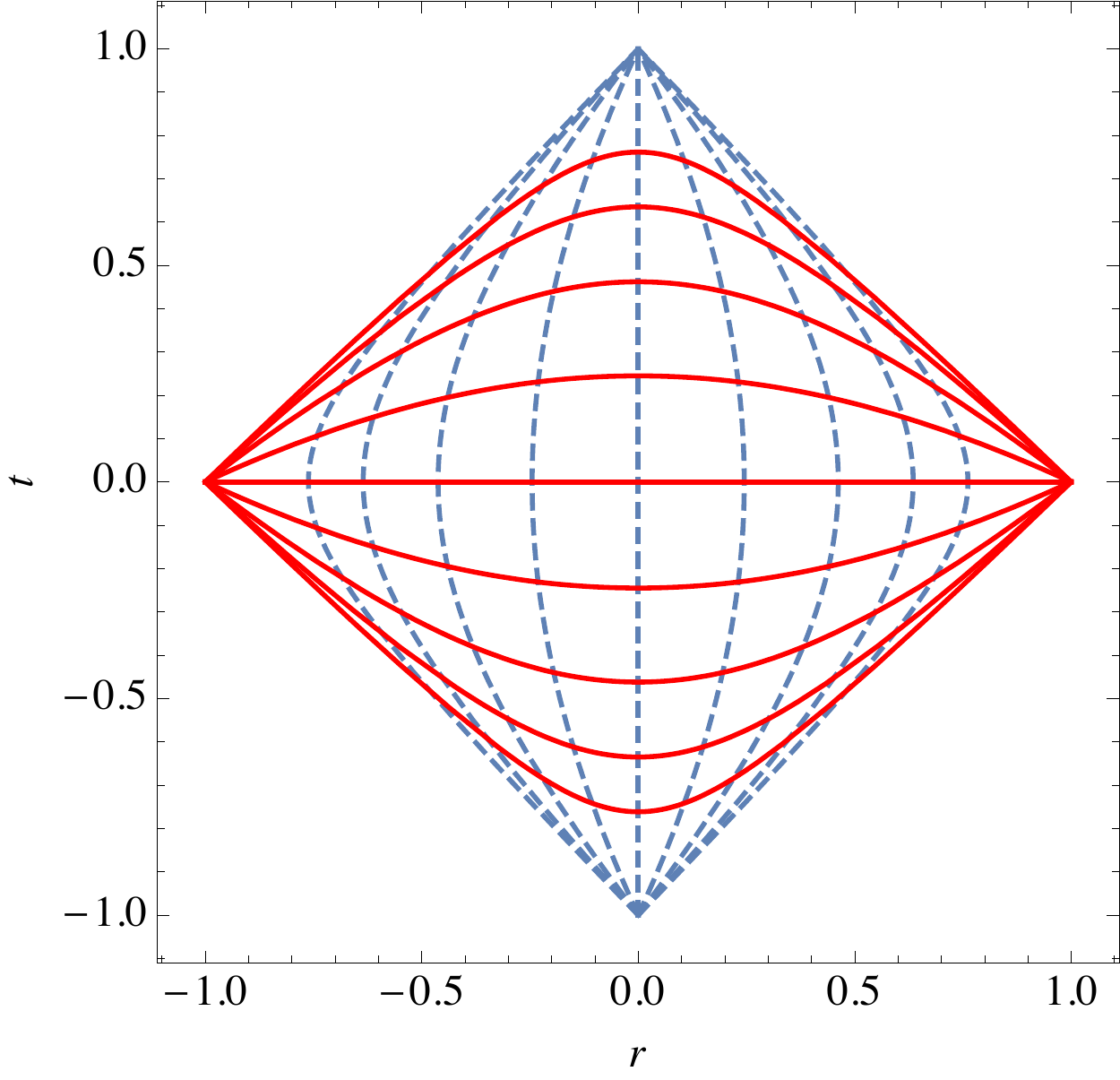}
b) \includegraphics[width = 2.5in]{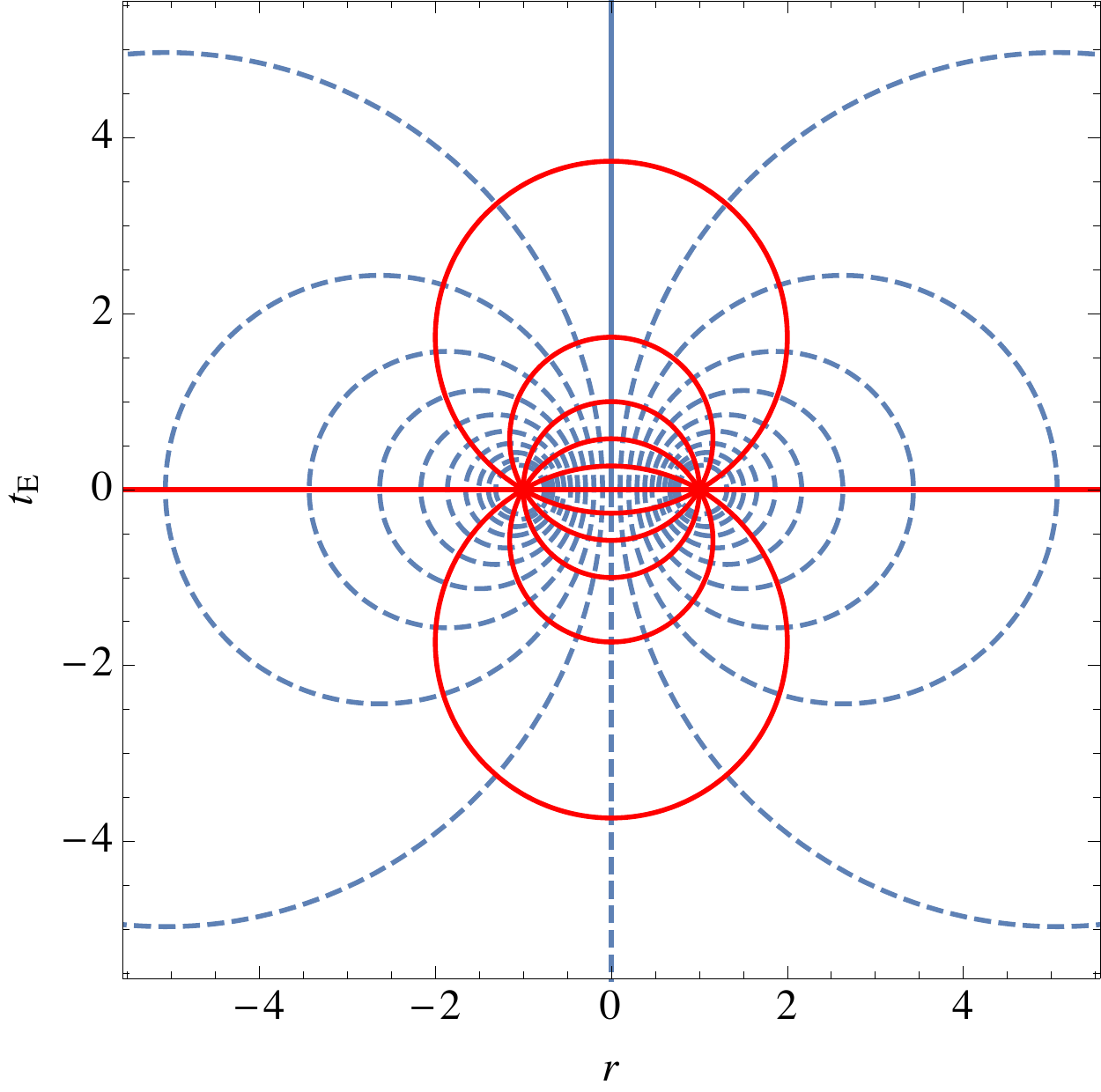}
\end{center}
\caption{a) Blue dashed curves are constant $u$ contours.  Red curves are constant $\tau$ contours.
b) Blue dashed curves are constant $u$ contours. Red curves are constant $\tau_E$ contours.
Note that we have plotted negative values for $r$ and $u$ even though both technically are restricted to be positive.
\label{contourplot}
}
\end{figure}

The computation of the entanglement entropy across a sphere thus reduces to a computation of the thermodynamic entropy of the hyperbolic space $S_E = 2 \pi \ell \langle H \rangle - W$ where $W \equiv - \Log \tr e^{-2 \pi \ell H}$.  As it did in $d=2$, this computation in turn breaks down into three pieces, a computation of $\langle H \rangle$, a computation of the effective anomaly action $\mathcal{W}[\delta_{\mu\nu}, e^{-2 \sigma} \delta_{\mu\nu}]$ and a computation of a universal contribution to $\widetilde W[\delta_{\mu\nu}]$,
\be
\label{E:SEhyperbolic}
S_E = 2\pi \ell \langle H\rangle + \W[\delta_{\mu\nu},e^{-2\sigma}\delta_{\mu\nu}] - \widetilde{W}[\delta_{\mu\nu}]\,.
\ee
To compute $\langle H \rangle$, we shall not try to write down the Schwarzian derivative in arbitrary even $d$, but instead rely on an earlier closely related computation performed in ref.\ \cite{Herzog:2013ed}.

We have not been able to compute $\mathcal{W}[\delta_{\mu\nu}, e^{-2 \sigma} \delta_{\mu\nu}]$ in general $d$, but we shall argue based on computations in $d=2$, 4 and 6 that it precisely cancels the contribution to $S_E$ from $\langle H \rangle$.  Finally, we compute $\widetilde W[\delta_{\mu\nu}]$ and show that the logarithmic contribution to it always reproduces the universal part of the sphere entanglement entropy.

\subsection{Casimir Energy}

The easy part of this computation is $\langle H \rangle$ because it has essentially been done in ref.\ \cite{Herzog:2013ed}. In that paper, two of us computed the stress tensor in the vacuum on $R \times S^{d-1}$ in even $d$, within the scheme where the the trace anomaly takes the form
\be
\langle T^{\mu}{}_{\mu}\rangle = \sum_j c_j I_j - (-1)^{\frac{d}{2}}\frac{4a}{d!\Vol(S^d)}E_d\,,
\ee
i.e. in a scheme where local counterterms are used to remove the total divergence from the stress tensor trace. Within that scheme, the stress tensor is unambiguously determined by $a$ to be
\be
\langle T^0_0 \rangle = - \frac{4 a }{(- \ell^2)^{d/2} d \Vol(S^d)} \ , \; \; \;
\langle T^i_j \rangle = \frac{4a}{(-\ell^2)^{d/2} d (d-1) \Vol(S^d)} \delta^i_j \ .
\ee
(Note the change in conventions for $a$ between that paper and this.) On $R \times H^{d-1}$ at the temperature $T = \frac{1}{2\pi \ell}$ it follows that
\be
\langle T^0_0 \rangle = -  \frac{4 a }{d \,  \ell^d  \Vol(S^d)}  \ , \; \; \;
\langle T^i_j \rangle =  \frac{4a}{d (d-1) \ell^d \Vol(S^d)} \delta^i_j \ ,
\ee
because the Riemann tensor is the opposite sign, and the result is constructed from the same product of $d/2$ Riemann tensors in each case. As the energy density is constant, the total energy is given by multiplying the energy density by the (divergent) volume of hyperbolic space, $\langle H \rangle = \langle T^{00} \rangle \Vol(H^{d-1})$. We need to isolate the logarithmic contribution to this volume
\be
\Vol(H^{d-1}) = \ell^{d-1} \Vol(S^{d-2}) \int_0^{u_{\rm max}} \sinh^{d-2} u \, \d u \ 
\ee
where our cut-off is 
\be
u_{\rm max} = - \Log \frac{ \delta / \ell}{2 - \delta / \ell} \ .
\ee
We find that
\be
\label{E:hyperbolicVolume}
\Vol(H^{d-1}) = \ldots +  \frac{(-1)^{d/2}}{\pi} \ell^{d-1} \Vol(S^{d-1}) \Log \frac{\delta}{\ell} + \ldots
\ee
and hence that
\be
\label{CFTH}
2 \pi \ell \langle H \rangle = \ldots + (-1)^{d/2} \frac{8 a }{ d } \frac{\Vol(S^{d-1})}{\Vol(S^d)} \Log \frac{\delta}{\ell} + \ldots
\ee

Like the stress tensor on $R\times S^{d-1}$, neither the stress tensor on $R\times H^{d-1}$ nor $\langle H\rangle$ is independent of the choice of scheme. For example, if one computes the partition function of a $d=4$ conformal field theory in two different schemes in $d=4$, their generating functionals may differ by the local counterterm
\be
\xi \int d^4x \sqrt{g} R^2\,,
\ee
where the coefficient $\xi $ is real. Taking a metric variation of the counterterm, it is clear that the stress tensor on $R\times S^{d-1}$, or $\langle H\rangle$ on $R\times H^{d-1}$, depends on the choice of $\xi$.  
See refs.\ \cite{Herzog:2013ed,Huang:2013lhw,Assel:2015nca} for lengthier discussions of this issue. 
However, the dependence of $W$ on $\xi$ is linear in $\beta$.  Thus
while $\langle H\rangle$ depends on the choice of scheme, the result we obtain for the sphere entanglement entropy $S_E$ does not. 

In principle, we should also worry about boundary contributions to $\langle H \rangle$.  We claim these contributions do not contribute to the logarithm.  One way to compute them is to look at the metric variation of the boundary $Q_{n,m}$ term in $n = d + \epsilon$ dimensions. As we saw before, the variation of the metric through the spin connection will cancel against a bulk variation of $E_{n,m}$.  The remaining variation comes only from the vielbeins, and cannot produce a logarithmic contribution.

\subsection{Dilaton Effective Action}

It is more involved to obtain $\mathcal{W}[\delta_{\mu\nu}, e^{-2 \sigma} \delta_{\mu\nu}]$. In $d=2, 4$, and $6$, we use the dilaton effective actions that we found in sections \ref{sec:twod} and \ref{sec:dilaton}. We will see that logarithmic contributions from $\langle H \rangle$ and $\mathcal{W}$ cancel out, i.e.\ that
\be
\label{nolog}
2 \pi \ell \langle H \rangle + \W[\delta_{\mu\nu}, e^{-2 \sigma} \delta_{\mu\nu}]
\ee 
has no logarithmic contribution. Thus,  
the entire entanglement entropy contribution comes from $\widetilde  W[\delta_{\mu\nu}]$, which we will compute next.

In principle we, should be able to evaluate $\W[\delta_{\mu\nu}, e^{-2 \sigma} \delta_{\mu\nu}]$ for general even $d$ and find the same cancelation of the logarithmic pieces.  In practice, there is an issue of non-commuting limits in dimensional regularization which makes the calculation difficult. The correct order of limits is to take the metric to be completely general, take the $n\to d$ limit, and only then specialize to the metric of interest. To see that the other order of limits is problematic, consider the following example.   
If we first fix the metric $e^{-2\sigma} \delta_{\mu\nu}$ to be that of $S^1\times H^{n-1}$ and then take the limit $n \to d$, we get a divergence that disappears in the other order of limits.  
Because $S^1 \times H^{n-1}$ contains an $S^1$ factor, the Euler characteristic, i.e.\ the leading $1/(n-d)$ singularity in $\widetilde W[e^{-2\sigma} \delta_{\mu\nu}]$, will vanish.  In contrast, the leading $1/(n-d)$ singularity from the boundary contribution to $\widetilde W[\delta_{\mu\nu}]$ will not vanish.  Thus $\W[\delta_{\mu\nu}, e^{-2\sigma} \delta_{\mu\nu}]$ computed in this order will not even be finite. 

We identify the conformal factor $\sigma$ in the metric (\ref{HSmetric}) with the dilaton $\tau$ of section \ref{sec:dilaton} (not to be confused with hyperbolic time). For convenience, we divide up the bulk and boundary contributions to $\mathcal{W}$.  We find the following results.

\subsubsection*{$d=2$}

The $d=2$ case can be evaluated from the effective action \eqref{Wefftwod}. Denoting $ {c\over 12}={a}$ and recalling that an interval has two endpoints, we find the bulk contribution to $\W$ is
\be
\W[\delta_{\mu\nu}, e^{-2 \sigma} \delta_{\mu\nu}]_{\rm{Bulk}}= -\left({a \over 2 \pi}\right)   \Big(2 \pi u - 4 \pi \Log(\sinh u)\Big) \Vol(S^0) +...
\ee
The boundary action contributes the following relevant divergence (the logarithmic divergence)
\be
\W[\delta_{\mu\nu}, e^{-2 \sigma} \delta_{\mu\nu}]_{\rm{Boundary}}=-\left({a \over 2 \pi} \right)  \Big(4 \pi u\Big) \Vol(S^0) +\hdots\,,
\ee 
so that the logarithmic contribution to $\W$ is
\be
\W[\delta_{\mu\nu},e^{-2\sigma}\delta_{\mu\nu}] = - 2au + \hdots\,.
\ee
Using the expression~\eqref{CFTH} for $\langle H\rangle$, we see that $2\pi\ell \langle H\rangle + \W[\delta_{\mu\nu},e^{-2\sigma}\delta_{\mu\nu}]$ has no logarithmic term.

\subsubsection*{$d=4$} 

In $d=4$, we find that the bulk and boundary terms in the expression \eqref{Weffdfour} for $\W$ contribute the following logarithmically divergent terms
\begin{align}
\begin{split}
\W[\delta_{\mu\nu}, e^{-2 \sigma} \delta_{\mu\nu}]_{\rm{Bulk}}&= \frac{ a}{(4 \pi )^2} \Big(6 \pi u-16 \pi \Log (\sinh u)\Big) {\rm{Vol}}(S^2)+\hdots \,,\\
\W[\delta_{\mu\nu}, e^{-2 \sigma} \delta_{\mu\nu}]_{\rm{Boundary}}&=\frac{ a}{(4 \pi )^2} \Big(16 \pi u\Big){\rm{Vol}}(S^2)+\hdots\,.
\end{split}
\end{align}

\subsubsection*{$d=6$} 

In $d=6$, we find that the bulk and boundary terms in the expression \eqref{6dresult} for $\W$ give
\begin{align}
\begin{split}
\W[\delta_{\mu\nu}, e^{-2 \sigma} \delta_{\mu\nu}]_{\rm{Bulk}}&=-\frac{ a}{(4 \pi )^3}  \left(30 \pi u- 96 \pi \Log(\sinh u) \right){\rm{Vol}}(S^4)+\hdots\,,\\
\W[\delta_{\mu\nu}, e^{-2 \sigma} \delta_{\mu\nu}]_{\rm{Boundary}}&=-\frac{ a}{(4 \pi )^3} (96 \pi u) {\rm{Vol}}(S^4)+\hdots\,.
\end{split}
\end{align}

In sum, using the dilaton effective action in $d=2,4,6$, we confirm that there is no logarithmic contribution to $2\pi \ell \langle H\rangle + \W[\delta_{\mu\nu},e^{-2\sigma}\delta_{\mu\nu}]$, as advertised.

\subsection{The Boundary Contribution to $W$ in General Dimension}

The last calculation to do is then an evaluation of the logarithmic contribution to $\widetilde W[\delta_{\mu\nu}]$ in general  dimension. To keep the boundary parametrization simple, it is useful to work in the $(\tau , u)$ coordinate system. In that system, we have that the extrinsic curvature takes the form
\be
K_\tau^\tau = - \frac{\sinh u}{\ell} \,, \qquad  K_u^u = 0 \,, \qquad
K_i^j = \frac{1}{\ell}(\cosh \frac{\tau}{\ell} \coth u + \csch u ) \delta^i_j \,.
\ee
The bulk term in $\widetilde{W}$ vanishes identically in flat space, so it remains to evaluate the boundary term. Two useful integrals for evaluating that boundary term in flat space are, for even $d$,
\begin{align}
\begin{split}
\int_0^{2\pi} \frac{(1 + \cosh u \cos t)^{d-2}}{(\cosh u + \cos t)^{d-1}} \d t &= \frac{\pi}{\sinh u} \frac{(d-2)!}{2^{d-3} \left( \frac{d-2}{2} ! \right)^2}  \ , \\
\int_0^1  (1-s^2)^{d/2-1}\d s &= \frac{\sqrt{\pi} \left(\frac{d-2}{2}\right)!}{2 \left( \frac{d-1}{2} \right)!} \ .
\end{split}
\end{align}
Starting with the expression (\ref{E:Qd}) and using the Gauss equation to replace the non-zero $\mathcal{\oR}_{\alpha\beta\gamma\delta}$ with the vanishing $R_{\mu\nu\rho\sigma}$, the logarithmic contribution to the boundary term is
\begin{eqnarray}
\label{Qdflat}
 \int_{\partial M} {\mathcal Q}_{n,d/2}
%
&=& \ldots + \frac{2 \pi (n-d) d!}{d-1} \Vol(S^{d-2}) \Log \frac{\delta}{\ell} + \ldots \ .
%
\end{eqnarray}
Using that for even $d$,
\be
\frac{\Vol(S^{d-2})}{\Vol(S^d)} = \frac{d-1}{2\pi}\,,
\ee
we then find the logarithmic contribution
\be
-\widetilde{W}[\delta_{\mu\nu}] = \ldots + (-1)^{d/2}4a \ln \frac{\delta}{\ell} + \ldots\,.
\ee
Using the expression~\eqref{E:SEhyperbolic} for $S_E$ and that $2\pi \ell \langle H\rangle + \W[\delta_{\mu\nu},e^{-2\sigma}\delta_{\mu\nu}]$ has no logarithmic term, we indeed find that the universal term in the entanglement entropy $S_E$ across a sphere is
\be
S_E = \ldots + (-1)^{d/2}4a \ln \frac{\delta}{\ell} +\ldots\,,
\ee
as claimed in ref.~\cite{Casini:2011kv}.

This computation is in fact almost a topological one. Under a constant rescaling $\sigma = \lambda$, the $a$-contribution to the Weyl anomaly guarantees that the generating functional $W$ varies on a manifold with Euler characteristic $\chi$ as (focusing just on the contribution proportional to $a$)
\be
\delta_{\lambda}W = (-1)^{d/2}(2a) \chi \lambda\,.
\ee
Now, the $4a$ in the entanglement entropy is essentially $(2a)\chi(S^{d-2}) $, and $\chi(S^{d-2})$ is the change in 
the Euler characteristic of flat space  
when a 
$D \times S^{d-2}$ is removed where $D$ is an open two dimensional disk. To see this, we use that the Euler characteristic satisfies an inclusion/exclusion principle $\chi(A \cup B) + \chi(A \cap B) = \chi(A) + \chi(B)$.  Let $A$ be ${\mathbb R}^d$ with a $D \times S^{d-2}$ removed.  Let $B = \overline D \times S^{d-2}$ be a closed set.  From the inclusion/exclusion principle, it follows that removing the $D \times S^{d-2}$ subtracts a $\chi(S^{d-2})$ from the Euler characteristic of the original space $A \cup B$. 

There is a sense in which introducing a boundary was not helpful.  Often in these types of computations, 
knowing the value of a difference like
$\W[\delta_{\mu\nu}, e^{-2 \sigma} \delta_{\mu\nu}]$ is useful because there are symmetry reasons to believe that for the reference background $\widetilde W[\delta_{\mu\nu}]$ will vanish.  Here, precisely because we had a boundary, $\widetilde W[\delta_{\mu\nu}]$ did not vanish.  As a result, we needed an independent way of calculating $\widetilde W[\delta_{\mu\nu}]$, and in fact, when the dust settled, we saw that we only needed to calculate $\widetilde W[\delta_{\mu\nu}]$.  Everything else canceled.

That $\widetilde W[\delta_{\mu\nu}]$ gives the right answer could perhaps have been anticipated.  From ref.\ \cite{Solodukhin:2008dh}, it is known at least in four dimensions that the $a$ dependent contribution to the entanglement entropy for a general entangling surface 
$\Sigma$ is 
proportional to the Euler characteristic of that surface, $S_E \sim 2 a \chi(\Sigma) \ln (\delta / \ell )$.  The fact that $\widetilde W[\delta_{\mu\nu}]$ gives us the entanglement entropy in our case could be viewed as confirmation of ref.\ \cite{Solodukhin:2008dh} in the case when $\Sigma$ is a sphere.  It is not too much of a stretch to imagine that in general even $d$, the $a$ dependent part of the entanglement entropy will be $S_E \sim (-1)^{d/2} 2 a \chi(\Sigma) (\ln \delta/  \ell)$.    Indeed, there are arguments to this effect in refs.\ \cite{Myers:2010tj,Myers:2010xs}.
That we are confirming in $d=4$ a specific case of a more general result is reassuring because 
evaluating $\widetilde W[\delta_{\mu\nu}]$ involves taking limits in a problematic order, as we already described above, first fixing the metric and then taking the number of dimensions $n \to d$.

Before proceeding, we write down an expression for the thermal partition function $W_H = - \ln Z_H$ on $H^{d-1}$ at temperature $T=1/(2\pi\ell)$ whose logarithmic pieces agree with the results above 
\be
W_H  = -a \frac{4\pi \ell }{(4 \pi \ell^2)^{d/2} \left(\frac{d}{2}\right)!} \left[  \Gamma(d) - 2^{d-1} \Gamma\left( 1+ \frac{d}{2} \right) \Gamma \left( \frac{d}{2} \right) \right] 
 \Vol(H^{d-1}) + \ldots \ .
 \label{Wholo}
\ee
The first term is proportional to $\langle H \rangle$ and the second term gives the entanglement entropy. The quantity in brackets is A160481 in the Online Encyclopedia of Integer Sequences \cite{oeis}.  

\subsection{A Different Conformal Transformation: de Sitter spacetime}

As we just saw, computing the entanglement entropy of a ball using the map to hyperbolic space is a rather intricate calculation that boils down, at the end of the day, to a computation in flat space of $\widetilde W[\delta_{\mu\nu}]$. In some sense, then, the conformal transformation is  unnecessary and does not give us extra information. We can try to see this phenomenon in a different example, a map from the causal development of the ball in flat space to the static patch of de Sitter spacetime. Ref.\ \cite{Casini:2011kv} already used this map in a successful calculation of the universal term in the entanglement entropy across a sphere, but we should revisit this computation in light of our boundary terms. In the Euclidean version of this map, the target space is an even-dimensional sphere $S^d$ with no boundary.  Naively, we can ignore boundary terms.  Nevertheless, the Weyl scale factor is not well behaved everywhere, and to be rigorous, we can introduce an artificial boundary to regulate its divergences.  
 
The metric $\eta$ on Minkowski space is related by a Weyl rescaling to the metric on the static patch of de Sitter   
\begin{align}
\begin{split}
\eta &=  -\d t^2 + \d r^2 + r^2 \d \Omega_{d-2}^2 \\
&= e^{2 \sigma} [ -\cos^2 \theta \d \tau^2 + \ell^2 (\d \theta^2 + \sin^2 \theta \, \d \Omega_{d-2}^2 ) ] \ , 
\end{split}
\end{align}
where
\be
\sigma  = - \ln (1 + \cos \theta \cosh(\tau/\ell) ) \ ,
\ee
and $0 < \theta < \pi/2$ while $-\infty < \tau < \infty$. The coordinates are related via the transformation
\begin{align}
\begin{split}
t &= \ell \frac{\cos \theta \sinh(\tau/\ell)}{1 + \cos \theta \cosh(\tau/\ell)} \ , \\
r &= \ell \frac{\sin \theta}{1 + \cos \theta \cosh(\tau/\ell)} \ .
\end{split}
\end{align}
The causal development of the ball, cut out by $\ell = \pm(t-r)$ and $\ell = \pm (t+r)$ is mapped to $e^{\pm \tau/\ell} = \tan \left(\frac{\theta}{2} - \frac{\pi}{4} \right)$.  In the Euclidean version of this map $\tau \to i \tau_E$, the boundary is reduced to the point $(\tau_E,\theta) = (0, \pi/2)$.  In contrast to the map to hyperbolic space where the boundary of the causal development mapped to the boundary of $H^{d-1}$, here the point $(0,\pi/2)$ is a smooth interior point of the $S^d$. The bulk integrals will not diverge here, and we do not need to introduce a regulated boundary.  

In contrast, the Weyl scaling factor $\sigma$ is divergent at the point $(\pi \ell, 0)$, and technically we should regulate the anomaly action by introducing a boundary here.  To do so, we introduce a local coordinate system in the vicinity of the point $(\pi \ell, 0)$, $\theta \approx \rho \sin \phi$ and $\tau_E/\ell - \pi \approx \rho \cos \phi$ where $0< \phi < \pi$ in order to keep $\theta >0$.  Near this point, the metric on flat space can be written
\be
\eta \approx \frac{4}{\rho^4} \left( \d \rho^2 + \rho^2 \d \phi^2 + \rho^2 \sin^2 \phi \, \d \Omega_{d-2}^2 \right) \ .
\ee
Introducing a boundary at $\rho = \delta \ll 1$, the nonzero components of the extrinsic curvature are  $K^\alpha_\beta = \frac{\rho}{2} \delta^\alpha_\beta$. It follows that
\begin{align}
\begin{split}
\int_{\partial M} \Q_d &= \left(\int_0^\pi \sin^{d-2} \d \phi\right) \left( \int_0^1 (1-s^2) \d s\right) \Vol(S^{d-2}) \, d! \, 2 \ln \delta \\
&= \frac{2 \pi d!}{d-1} \ln \delta \ .
\end{split}
\end{align}  
It is then straightforward to see that the logarithmic contribution to $\widetilde W[\delta_{\mu\nu}]$ and the boundary logarithmic contribution to $\W[\delta_{\mu\nu}, e^{-2 \sigma} \delta_{\mu\nu}]$ are identical.  Moreover, these logarithmic contributions are the same as was found using a different boundary and the map to hyperbolic space.  This equivalence is not surprising since the contributions are topological in nature, and the boundaries, though different, are still topologically the same. 
 
As already mentioned in ref.\ \cite{Casini:2011kv}, the Casimir energy in de Sitter spacetime does not contribute to the logarithmic divergence and the full logarithmic term of the entropy is dictated by the partition function evaluated in the curved metric. The expression $\W[ \delta_{\mu\nu}, e^{-2 \sigma} \delta_{\mu\nu}]$ has bulk contributions from de Sitter and from the ball but also, now in light of our results, a boundary contribution from the surface $\rho = \delta$.  Ref.\ \cite{Casini:2011kv} got the right answer purely from the bulk contribution to $\W[ \delta_{\mu\nu}, e^{-2 \sigma} \delta_{\mu\nu}]$. As follows from the previous paragraph, had they computed the boundary contribution as well, they would have found, like us, that $2 \pi \ell \langle H \rangle + \W[ \delta_{\mu\nu}, e^{-2 \sigma} \delta_{\mu\nu}]$ has no logarithmic contribution and that the entire log contribution can be attributed to $\widetilde W[\delta_{\mu\nu}]$. For example, using our explicit anomaly action in $d=4$, we find
\be
\W[\delta_{\mu\nu}, e^{-2 \sigma} \delta_{\mu\nu}]|_{\rm Bulk} \sim \frac{a}{(4 \pi)^2} 16 \pi \ln \delta \Vol(S^2)  = 4 a \ln \delta\,,
\ee
where we integrate only from $- \pi \ell + \delta < \tau_E < \pi \ell - \delta$.  

Interestingly, though, the bulk contribution to $\W[\delta_{\mu\nu},e^{-2\sigma}\delta_{\mu\nu}]$ considered in ref.\ \cite{Casini:2011kv} did give the correct answer for the entanglement entropy on its own. Similarly, in our case of the map to hyperbolic space, we could have thrown out the equal and opposite contributions from $\W[ \delta_{\mu\nu}, e^{-2 \sigma} \delta_{\mu\nu}]|_{\rm Boundary}$ and $\widetilde W[\delta_{\mu\nu}]$ and also gotten the correct answer purely from $\W[ \delta_{\mu\nu}, e^{-2 \sigma} \delta_{\mu\nu}]|_{\rm Bulk}$. As the split between bulk and boundary terms in $\W[ \delta_{\mu\nu}, e^{-2 \sigma} \delta_{\mu\nu}]$ is arbitrary up to a choice of which total derivatives to include in the bulk action, getting the correct answer from $\W[ \delta_{\mu\nu}, e^{-2 \sigma} \delta_{\mu\nu}]|_{\rm Bulk}$ alone appears to be a coincidence. In fact, at least regarding logarithmic terms, we have specified a separation between bulk and boundary terms by insisting that the only place in which $\tau$ appears without a derivative in the boundary action is multiplying $Q_d$.  This split has the advantage of giving the boundary contribution a topological interpretation when the reference metric is flat.  Indeed, given this choice,
it becomes manifest for the two maps we considered 
that both $\W[ \delta_{\mu\nu}, e^{-2 \sigma} \delta_{\mu\nu}]|_{\rm Boundary}$ and 
$\widetilde W[\delta_{\mu\nu}]$ will yield the Euler characteristic of the flat space multiplied by a logarithm of the UV cut-off.

\section{Discussion}
\label{sec:discussion}

We resolved the puzzle described in ref.\ \cite{Casini:2011kv}: the universal logarithmic term in the entanglement entropy~\eqref{SEball} across a sphere in flat space (for a conformal theory) can be recovered by a Weyl transformation to hyperbolic space, provided one keeps careful track of boundary terms. One interesting consequence of our results is that the logarithmic term can be interpreted as a purely boundary effect. With the help of the conformal map to  hyperbolic space cross a circle, focusing on the universal part, we identify the logarithmic contribution to the entanglement entropy $S_E$ and the dimensionally regularized effective action $\widetilde W[\delta_{\mu\nu}]$: 
\be
S_{\rm E}\equiv - \tr (\rho_A \ln \rho_A) \sim -\widetilde W[\delta_{\mu\nu}] \ ,
\ee
where $\widetilde W[\delta_{\mu\nu}]$ is given by eq.~\eqref{tildeWgeneral}. 
$\widetilde W[\delta_{\mu\nu}]$ corresponds to a dimensionally continued Euler characteristic of the causal development of the interior of the sphere, a ball, which in turn receives contributions purely from the spherical boundary of the ball since the Riemann curvature and hence the Euler density vanish in flat space. The leading area law divergence in the entanglement entropy is also usually interpreted to be a boundary effect: entanglement entropy scales with the area of the boundary because in the ground state most of the entanglement is assumed to be local. But here we see that the subleading logarithmic divergence is also a boundary effect. Perhaps this result  should have been anticipated since both divergences are regulated by a short distance cut-off $\delta$, which one could think of as the distance between lattice points on either side of the boundary.

As we discussed in section \ref{sec:example}, that $\widetilde W[\delta_{\mu\nu}]$ on its own gives the correct answer for the log term in the entanglement entropy across a sphere can be viewed as a special case of Solodukhin's result \cite{Solodukhin:2008dh} using a squashed cone in $d=4$ that the $a$ contribution to the entanglement entropy across a general surface $\Sigma$ can be written
\be
S_E \sim 2 a \chi(\Sigma) \ln (\delta / \ell) \ .
\ee
For non-spherical entangling surfaces, there will of course be other contributions to $S_E$, for example from the $c_j$ central charges.  
While we are not aware of a derivation (refs. \cite{Myers:2010tj,Myers:2010xs} come close but ultimately only consider the sphere case), it seems reasonable that in general dimension, the only modification needed to make this formula correct in our conventions is a factor of $(-1)^{d/2}$.  

In the process of resolving this puzzle, we produced a number of auxiliary results which are interesting in their own right. In two dimensions, where the trace anomaly is perhaps most powerful, we were able to use an effective anomaly action to reproduce three well-known results in conformal field theory, namely the Schwarzian derivative, the entanglement entropy of an interval, and also the R\'enyi entropies for the interval.  Neither the effective anomaly action we use nor the results are new. However, we have not seen our form of the effective anomaly action used to derive these three results before.\footnote{%
 See however ref.\ \cite{Solodukhin:1994yz} for a similar calculation.
}
Additionally, the story in two dimensions provides a simple warm-up example for the story in general dimension which we pursued next.

Between $d=4$ and $d=6$, our story is the most complete in $d=4$. 
%
In four dimensions, we derived from general principles the most general Wess-Zumino consistent result for the trace anomaly on a manifold with a codimension one boundary, including two boundary central charges we denoted $b_1$ and $b_2$.  
It would be interesting to study $b_1$ and $b_2$ further (as well as their counter-parts in higher dimensions). 
What values\footnote{%
Shortly after the first version of this paper appeared on the arXiv, these boundary central charges for fermions and gauge fields were computed in $d=4$ in ref.\ \cite{Fursaev:2015wpa}.
}
 do they take for massless fermions?  for a gauge field?  for superconformal field theories? Might they be ordered under renormalization group flows, like the coefficient $a$?  

Another pair of key results in this paper are explicit formulae with boundary terms for the $a$ contribution to the effective anomaly action in $d=4$ and $d=6$ dimensions.  Previously, to our knowledge, only the bulk contribution had been worked out \cite{Komargodski:2011vj, BrownOttewill, Elvang:2012st}. Unfortunately, in $d=6$, we were only able to detail the boundary contribution to the action for a conformally flat metric. The conformally flat case was enough to study the entanglement entropy across a sphere. Nevertheless, it would be nice to write down the boundary contribution for a general metric.  

It would also of course be interesting to see if the $a$ contribution to the effective anomaly action can be given an explicit and simple form in any dimension. That the sphere entanglement comes solely from $\widetilde  W[\delta_{\mu\nu}]$ depended on cancellation between the Casimir energy $\langle H \rangle$ and the effective anomaly action $\W[\delta_{\mu\nu}, e^{-2 \sigma} \delta_{\mu\nu}]$ that we were only able to verify explicitly in $d=2$, 4 and 6.  In general even dimension, we were hampered by non-commuting limits that forced us to fix $d$ before choosing a metric in order to calculate $\W[\delta_{\mu\nu}, e^{-2 \sigma} \delta_{\mu\nu}]$.

In appendix~\ref{app:holo} we reproduce the holographic computation of the sphere entanglement entropy using hyperbolic space.  Holographic renormalization allows us to write down a regulated effective action $W_H$ for $S^1 \times H^{d-1}$ itself without need for a reference background.  Thus we are saved the trouble that we faced with our dilaton effective action 
of needing to compute $W$ for the reference background.  
 
Another interesting result of the holographic calculation is the vanishing of the second derivative of the effective action $W_H$~\eqref{holosecderiv}. While experience suggests that the result is the consequence of a Maxwell relation combined with scale invariance, we have not been able to prove the vanishing for a general conformal field theory. 

Finally, in this paper we mostly adopted the dimensional regularization to construct $\W$. It would be interesting to construct $\W$ using the integral formula (\ref{E:integratedAnomalyGeneral}). 

\vskip 0.1in

\section*{Acknowledgments}
We would like to thank K.~Balasubramanian, H.~Casini, A.~O'Bannon, M.~Ro\v{c}ek,  W.~Siegel, 
and P.~van~Nieuwenhuizen for useful discussion. This work was supported in part by the NSF under Grant No.\ PHY13-16617. 

\appendix

\section{Differential Geometry with a Boundary}
\label{app:diffgeom}

Let $M$ be a $d$-dimensional, orientable, Riemannian manifold with metric $g$. 
In general $M$ will have a boundary $\partial M$. We use $x^{\mu}$ to indicate coordinates on patches of $M$ and $y^\alpha$ for coordinates on patches of $\partial M$. The boundary can be specified by means of the embedding functions $X^{\mu}(y^\alpha)$. These do not transform as tensors under reparameterizations in $M$, but their derivatives
\beq
f_\alpha{}^{\mu} \equiv \partial_\alpha X^{\mu}\,,
\eeq
do. Rather, the $f_\alpha^{\mu}$ transform as a vector under reparameterizations of the $x^{\mu}$ and as a one-form under reparameterizations of the $y^\alpha$. The $f_\alpha^{\mu}$ allow us to pull back covariant tensors on $M$ to covariant tensors on $\partial M$. For instance, the metric $g$ pulls back to the induced metric $\gamma$ with components
\beq
\og_{\alpha\beta} (y)= f_\alpha{}^{\mu}(y)f_\beta{}^{\nu}(y)g_{\mu\nu}(X(y))\,.
\eeq
We also define
\beq
f^\alpha{}_{\mu} \equiv g_{\mu\nu} \gamma^{\alpha\beta} f_\beta{}^{\nu} \,,
\eeq
which satisfies
\beq
f^\alpha{}_{\mu} f_\beta{}^{\mu} = \delta^\alpha_\beta\,, \qquad f^\alpha{}_{\mu}f_\alpha{}^{\nu} \equiv h^{\nu}_{\mu}\,,
\eeq
with $h^{\mu\nu}$ a tangential projector. We can also define a unit-length vector field $n^{\mu}$ after picking an orientation on $\partial M$ via
\beq
n^{\mu} =\frac{1}{(d-1)!} \varepsilon^{\mu}{}_{ \nu_1\hdots \nu_{d-1}} \varepsilon^{\alpha_1\hdots \alpha_{d-1}} f_{\alpha_1}{}^{\nu_1}\hdots f_{\alpha_{d-1}}{}^{\nu_{d-1}}\,.
\eeq
Throughout we take the orientation on $\partial M$ to be such that $n^{\mu}$ is always pointing outward.

\subsection{The Covariant Derivative and the Second Fundamental Form}

We use the Levi-Civita connection built from $g$ to take derivatives ${\rm D}$ on $M$. From this connection we construct a connection on $\partial M$ that allows us to take derivatives $\oD$ of tensors on $\partial M$. $\oD$ acts on e.g. a mixed tensor $\mathfrak{T}^{\mu}{}_\alpha$ via
\beq
\oD_\alpha \mathfrak{T}^{\mu}{}_\beta = \partial_\alpha \mathfrak{T}^{\mu}{}_\beta + \Gamma^{\mu}{}_{\nu \alpha}\mathfrak{T}^{\nu}{}_\beta - \oG^{\gamma}{}_{\beta\alpha} \mathfrak{T}^{\mu}{}_\gamma\,,
\eeq
with
\beq
\Gamma^{\mu}{}_{\nu \alpha} = \Gamma^{\mu}{}_{\nu\rho} f_\alpha{}^{\rho} \,, \qquad 
\oG^\alpha{}_{\beta\gamma}= f^\alpha{}_{\mu} \left(\partial_\gamma \delta^{\mu}_{\nu} + \Gamma^{\mu}{}_{\nu c}\right)f_\beta{}^{\nu}\,.
\eeq
It is easy to show that $\oG^\alpha{}_{\beta\gamma}$ is the Levi-Civita connection constructed from the induced metric $\gamma_{\alpha\beta}$, and furthermore that the derivative satisfies
\beq
\oD_\alpha g_{\mu\nu} = 0\,, \qquad \oD_\alpha \gamma_{\beta \gamma} = 0\,.
\eeq

There is a single tensor with one derivative that can be built from the data at hand, namely the second fundamental form $\II^{\mu}{}_{\alpha\beta}$,
\beq
\II^{\mu}{}_{\alpha\beta} \equiv \oD_\alpha f_\beta{}^{\mu}\,.
\eeq
One can show that
\beq
\II^{\mu}{}_{\alpha\beta} = \II^{\mu}{}_{\beta\alpha}\,, \qquad h_{\mu\nu} \II^{\nu}{}_{\alpha\beta}=0\,,
\eeq
and the latter implies that
\beq
\II^{\mu}{}_{\alpha\beta} = -n^{\mu} K_{\alpha\beta}\,,
\eeq
where $K_{\alpha\beta}$ is the extrinsic curvature of the boundary. From this and $n_{\mu}\oD_\alpha n^{\mu}=0$ we also find
\beq
\oD_\alpha n_{\mu} = f^\beta{}_{\mu} K_{\alpha\beta}\,.
\eeq

Let us relate this presentation to the more common one in terms of Gaussian normal coordinates. For some patch on $M$ which includes a patch of $\partial M$, we choose coordinates so that $g$ takes the form
\beq
g = \d r^2 + \hat{g}_{\alpha\beta}(r,y) \d y^\alpha \d y^\beta\,,
\eeq
where the boundary is extended in the $y^\alpha$ at $r=0$. That is, the embedding functions are $f_\alpha{}^r = 0\,, f_\alpha{}^\beta = \delta_\alpha^\beta$, and consequently the induced metric is 
\beq
\gamma_{\alpha \beta}(y) = \hat{g}_{\alpha \beta}(r=0,y)\,.
\eeq
In this coordinate choice we have
\beq
n^r = 1\,, \qquad \II^r{}_{\alpha \beta} = \Gamma^r{}_{\alpha \beta} = - \frac{1}{2}\left.\partial_r \hat{g}_{\alpha \beta}\right|_{y=0}\,.
\eeq
Note that the trace of the extrinsic curvature, $K= \gamma^{\alpha\beta}K_{\alpha \beta}$ is
\beq
K = \frac{1}{2}\left. \hat{g}^{\alpha \beta} \partial_r \hat{g}_{\alpha \beta}\right|_{r=0} =  \left.\frac{\pounds_n \sqrt{\hat{g}}}{\sqrt{\hat{g}}}\right|_{r=0}\,,
\eeq
with $\pounds_n$ the Lie derivative along $n^{\mu}$, which coincides with a common formula used by physicists for the extrinsic curvature of a spacelike boundary. 

\subsection{Gauss and Codazzi}

Consider the Levi-Civita connection one-form $\Gamma^{\mu}{}_{\nu} = \Gamma^{\mu}{}_{\nu\rho}\d x^{\rho}$ and its curvature
\beq
R^{\mu}{}_{\nu} = \d\Gamma^{\mu}{}_{\nu} + \Gamma^{\mu}{}_{\rho}\wedge \Gamma^{\rho}{}_{\nu} = \frac{1}{2}R^{\mu}{}_{\nu\rho\sigma}\d x^{\rho}\wedge \d x^{\sigma}\,.
\eeq
Here $R^{\mu}{}_{\nu\rho\sigma}$ is the Riemann curvature which can also be defined through the commutator of derivatives
\beq
[{\rm D}_{\rho},{\rm D}_{\sigma}]\mathfrak{v}^{\mu} = R^{\mu}{}_{\nu\rho\sigma}\mathfrak{v}^{\nu}\,,
\eeq
for $\mathfrak{v}^{\mu}$ a vector field. The pullback of $R^{\mu}{}_{\nu}$ to $\partial M$ can be expressed in terms of the curvature $\oR^{\mu}{}_{\nu}$ of $\oG$ and the second fundamental form. The resulting expressions are the Gauss and Codazzi equations. They can be summarized as
\beq
\text{P}[R^{\mu}{}_{\nu}] = \oR^\alpha{}_\beta f_\alpha{}^{\mu} f^\beta{}_{\nu} + \oD \mathcal{M}^{\mu}{}_{\nu}- \mathcal{M}^{\mu}{}_{\rho} \wedge \mathcal{M}^{\rho}{}_{\nu}\,,
\eeq
where $\oD$ is the covariant exterior derivative and
\beq
\mathcal{M}^{\mu}{}_{\nu} = \II^{\mu}{}_\alpha f^\alpha{}_{\nu} - f_\alpha{}^\mu \II_{\nu}{}^\alpha\,, \qquad \II^{\mu}{}_\alpha \equiv \II^{\mu}{}_{\alpha\beta} \d y^\beta\,.
\eeq
Alternatively, we can define
\beq
\tG^{\mu}{}_{\nu} = \Gamma^{\mu}{}_{\nu \alpha} \d y^\alpha - \mathcal{M}^{\mu}{}_{\nu}\,,
\eeq
whose curvature satisfies
\beq
\tR^{\mu}{}_{\nu} = \oR^\alpha{}_\beta f_\alpha{}^{\mu} f^\beta{}_{\nu}\,.
\eeq

In components, the Gauss and Codazzi equations read
\begin{align}
\begin{split}
\label{E:gaussCodazzi}
R_{\alpha \beta \gamma \delta} &= \oR_{\alpha \beta \gamma \delta} - K_{\alpha \gamma}K_{\beta \delta} + K_{\alpha \delta }K_{\beta \gamma}\,,
\\
R_{\mu \alpha \beta \gamma}n^{\mu} & = -\oD_\beta K_{\alpha \gamma} + \oD_\gamma  K_{\alpha \beta}\,,
\end{split}
\end{align}
and we have used the embedding scalars to convert indices on the bulk Riemann tensor into indices on $\partial M$.

\section{Wess-Zumino Consistency in $d=4$}
\label{A:WZ4d}

We now perform the algorithm described in Subsection~\ref{S:classify4d}, beginning with step 1. We need to parameterize the most general variation of $W$, which we denote as $\delta_{\sigma} W_b$. After some computation, we find that this variation contains sixteen independent terms\footnote{In compiling the list of these sixteen terms, we have made extensive use of the Gauss and Codazzi equations~\eqref{E:GaussCodazzi2}. We also use that the action of $n^{\mu}{\rm D}_{\mu}$ is only well-defined on bulk tensors.}
\beq
\delta_{\sigma}W_b = \int_{\partial M}\d^3y \sqrt{\gamma}\left\{ \sum_{I=1}^{8} b_I \B_I  + \sum_{J=1}^8 B_J\D_J \right\}  \delta\sigma\,,
\eeq
indexed by the eight $b_I$ and eight $B_J$. (The coefficients $b_I$ and $B_J$ are used to denote boundary central charges.) We organize the terms in the following way. The eight $\B_I$ are three-derivative scalars. The eight $\D_J$ all involve derivatives of the Weyl variation $\delta \sigma$, and so we denote them with a calligraphic $\D$ to suggest a derivative. 
We distinguish the $\B_I$ and $\D_J$ for two reasons. First, the allowed three-derivative counterterms are given by the $\B_I$. Second, we will see shortly that those local counterterms redefine the coefficients of the $\D_J$.

In any case, the $\B_I$ are
\begin{align}
\begin{split}
	\B_1 &= \oR K\,, 
	\quad 
	\B_2 = R K\,, 
	\quad
	\B_3 = \oR_{\alpha\beta}K^{\alpha\beta}\,, 
	\quad
	\B_4 = \tr K^3\,, 
	\\
	\B_5 &= K^3\,, 
	\quad 
	\B_6 =n^{\mu}\partial_{\mu} R  \,,
	\quad
	\B_7 = \tr \hat{K}^3\,, 
	\quad 
	\B_8 =W_{\alpha\beta\gamma\delta}\gamma^{\alpha\gamma}\hat{K}^{\beta\delta}\,, 
\end{split}
\end{align}
Here $W_{\alpha\beta\gamma\delta}$ is the pullback of the Weyl tensor to the boundary, and we have defined $\hat{K}$ to be the traceless part of the extrinsic curvature,
\be
\hat{K}_{\alpha\beta} \equiv K_{\alpha\beta} - \frac{K}{d-1}\gamma_{\alpha\beta}\,,
\ee
which transforms covariantly under Weyl rescaling as $\hat{K}_{\alpha\beta} \to e^{\sigma}\hat{K}_{\alpha\beta}$. $\B_7$ and $\B_8$ are then manifestly covariant under Weyl rescaling. They are the only nonzero scalars that can be formed from either three factors of $\hat{K}$, or one factor of $\hat{K}$ and one of the Weyl tensor. They cannot be eliminated by the addition of a local counterterm and are trivially Wess-Zumino consistent, and so represent genuine boundary anomalies. The $\text{tr}(\hat{K}^3)$ term first appeared in ref.\ \cite{Melmed}, while the $W_{\alpha\beta\gamma\delta}\gamma^{\alpha\gamma}\hat{K}^{\beta\delta}$ term appeared later in ref.\ \cite{Dowker:1989ue}. The $\D_J$ are
\begin{align}
\begin{split}
	\D_{1} &=\mathring{\Box}K \,, 
	\quad  
	\D_{2} =\oD_{\alpha}\oD_{\beta}K^{\alpha\beta} \,, 
	\quad 
	\D_{3}= \oR n^{\mu}\partial_{\mu}\,, 
	\quad 
	\D_{4} = R n^{\mu}\partial_{\mu}\,.
	\\
 	\D_{5}& = K_{\alpha\beta}K^{\alpha\beta} n^{\mu}\partial_{\mu} \,, 
	\quad 
	\D_{6} = K^2 n^{\mu}\partial_{\mu}\,, 
	\quad 
	\D_{7} =K n^{\mu}n^{\nu}{\rm D}_{\mu}{\rm D}_{\nu}\,,  
	\quad 
	\D_{8} = n^{\mu}n^{\nu}n^{\rho}{\rm D}_{\mu}{\rm D}_{\nu}{\rm D}_{\rho}\,, 
\end{split}
\end{align}

Continuing with step 2, the most general local boundary counterterm is
\beq
W_{CT} = \int_{\partial M}\d^3y \sqrt{\gamma}\sum_{I=1}^{6} d_I \B_I\,.
\eeq
The $d_I$ represent a choice of scheme. They can be adjusted to eliminate various coefficients in $\delta_{\sigma}W_b$. We would like to deduce which coefficients can be eliminated. This is an exercise in linear algebra. As $\sqrt{\gamma} \B_7$ and $\sqrt{\gamma} \B_8$ are invariant under Weyl rescalings, 
we do not include them in $W_{CT}$. 
The Weyl variation of $W_{CT}$ may then be understood as a linear map $\Sigma:\mathbb{R}^6\to\mathbb{R}^8$ which maps the $\{ \B_I \}$ (for $I=1,..,6$) to the $\{ \D_J\}$ as
\beq
\delta_{\sigma} \int_{\partial M} \d^3y\sqrt{\gamma} \,\B_I = \int_{\partial M} \d^3y \sqrt{\gamma}\sum_{J=1}^8 \Sigma^J{}_I \D_J\,.
\eeq
The number of $\D_J$ which can be eliminated is given by the dimension of the image of $\Sigma$, and the null vectors of $\Sigma^t$ encode the linear combinations of the $\D_J$ which cannot be removed by a judicious choice of scheme.

A straightforward computation gives
\beq
\label{E:sigma}
\Sigma = 
	\begin{pmatrix} 
		 -4 & -6 & -1  & 0 & 0 & 6 
		\\  
		 0 & 0 & -1  & 0 & 0 & 0 
		\\ 
		 3 & 0 & 1  & 0 & 0 & -3 
		\\
		 0 &3 & 0 & 0 & 0 & 1
		\\
		 0 & 0 & 0  & 3 & 0 & 3
		\\
		 0 & -6 &  0  & 0 & 9 & 3
		\\
		 0 & -6 &  0  & 0 & 0 & -6
		\\
		 0 & 0 & 0 & 0 & 0 & -6
	\end{pmatrix}
\eeq
The map $\Sigma$ is injective,
so six of $\D_J$ can be eliminated. The null vectors of $\Sigma^t$ are given by
\beq
\chi_1 = \begin{pmatrix} 3 & 1 & 4 & 0 & 0 & 0 & -3 & 4 \end{pmatrix} \,, \qquad \chi_2 = \begin{pmatrix} 0 & 0 & 0 & 6 & 0 & 0 & 3 & -2 \end{pmatrix}\,,
\eeq
so the image of $\Sigma$ is given by $\mathbb{R}^8$ modulo the $\mathbb{R}^2$ spanned by $\chi_1$ and $\chi_2$. In terms of the $\D_J$, the linear combinations
\beq
3\D_1 +\D_{2} + 4 \D_{3} -3\D_7 + 4 \D_8\,, \qquad 6 \D_{4} + 3\D_{7}-2\D_8\,,
\eeq
are never generated from the variation of $W_{CT}$. Said another way, the $d_I$ can be adjusted to eliminate all of the $\D_J$ except for $\D_1$ and $\D_4$. So the most general boundary Weyl variation, having modded out by local counterterms, is
\beq
\delta_{\sigma}W_b = \int_{\partial M}\d^3y\sqrt{\gamma}\left\{ \sum_{I=1}^{8} b_I \B_I + B_1 \mathring{\Box}K +B_4 Rn^{\mu}\partial_{\mu}   \right\} \delta \sigma\,.
\eeq

Now we implement step 3, by computing the second Weyl variation. The second variations of $\mathcal{B}_1 \delta \sigma_2$ through $\mathcal{B}_8 \delta \sigma_2$ follow (almost)  immediately from the $\delta_\sigma W_{CT}$ that we computed above.  Let us then consider carefully the second Weyl variation of the terms proportional to $B_1$ and $B_4$.  From these terms we get
\begin{eqnarray}
\delta_{\sigma_1}\delta_{\sigma_2} W_b &=& \int \d^3y \sqrt{\gamma} \left\{
B_1
\left( 3 (n^\mu \partial_\mu \delta \sigma_1) (\oBox \delta \sigma_2) + 2 K (\partial^\alpha \delta \sigma_1) (\partial_\alpha \delta \sigma_2) \right)
\right. \nonumber \\
&& \qquad\qquad  \left.  -6 B_{4} ( n^{\mu}\partial_{\mu} \delta \sigma_2) \left( \mathring{\Box}+ n^{\nu}n^{\rho}{\rm D}_{\nu}{\rm D}_{\rho} + K n^{\nu}\partial_{\nu}\right)\delta \sigma_1 + 
\ldots \right\} \ , 
\end{eqnarray}
where the ellipsis denotes terms that depend on $b_1$ through $b_6$. The only terms with a normal derivative of $\delta \sigma_2$  come from $B_4$.  Given that fact, it is impossible to symmetrize under $\delta \sigma_1\leftrightarrow \delta \sigma_2$ the term involving one normal derivative of $\delta \sigma_2$ and two normal derivatives of $\delta \sigma_1$. Thus Wess-Zumino consistency forces $B_{4} = 0$.

It is slightly more involved to see that $B_1$ must vanish. 
First, observe that the $\mathcal{B}_6$ term is the only one which produces a second variation $\delta \sigma_2 \mathcal{D}_8 \delta \sigma_1$, which has three normal derivatives and is not symmetric under $\delta\sigma_1\leftrightarrow\delta\sigma_2$ and so is not WZ consistent. So $b_6=0$. In fact, the same sort of reasoning tells us that $b_2=b_4=b_5=0$ and that $b_3$ is proportional to $b_1$ as $b_3 = -3 b_1$. In terms of the remaining parameters $b_1,B_1$, the second Weyl variation is simply
\be
\delta_{\sigma_1}\delta_{\sigma_2}W_b = \int {\rm d}^3y \sqrt{\gamma} \left\{3 b_1 \delta \sigma_2\hat{K}^{\alpha\beta} \oD_{\alpha}\oD_{\beta}\delta \sigma_1 + B_1 \left( 3 (n^{\mu}\partial_{\mu}\delta \sigma_1)(\oBox \delta \sigma_2) + 2 K(\partial^{\alpha}\delta\sigma_1)(\partial_{\alpha}\delta\sigma_2)\right)\right\}\,.
\ee
This expression is not symmetric under $\delta\sigma_1 \leftrightarrow \delta \sigma_2$ for any nonzero value of $b_1$ and $B_1$, and so WZ consistency enforces that they both vanish $b_1=B_1=0$.


The only ``boundary central charges'' that survive are $b_7$ and $b_8$, and the boundary term in the anomaly is
\be
\delta_{\sigma}W_b = \int_{\partial M}\d^3y\sqrt{\gamma} \left\{ b_7 \text{tr}\hat{K}^3 + b_8 \gamma^{\alpha\gamma}\hat{K}^{\beta\delta}W_{\alpha\beta\gamma\delta}\right\}\,.
\ee
Putting the pieces together, the total anomaly is given by~\eqref{E:4dTotalAnomaly} as advertised in Subsection~\ref{S:classify4d}.  In the text, we relabel: $b_7 \to b_1$ and $b_8 \to b_2$.  

\section{Effective Action from Dimensional Regularization}
\label{app:dimreg}

In this appendix we consider the anomaly effective action $\W$ in even $d$ dimensions as obtained from dimensional regularization via the expression~\eqref{E:dWgeneral}, which we recall here
\be
\label{E:dWgeneralv2}
\W[g_{\mu\nu},e^{-2\tau}g_{\mu\nu}] = A\lim_{n\to d} \frac{1}{n-d}\left\{\left(  \int_M \mathcal{E}_{n,m} - \int_{\partial M} \mathcal{Q}_{n,m} \right)-\left( \int_M \hat{\mathcal{E}}_{n,m} - \int_{\partial M}\hat{\mathcal{Q}}_{n,m}\right)\right\}\,, 
\ee
where $m=d/2$ and $A = (-1)^{d/2}4a/(d!\Vol(S^d))$. Here we obtain the explicit forms of $\W$ in $d=4,6$ including boundary terms.  (In $d=6$ the boundary action will be evaluated in a conformally flat geometry.) The bulk dilaton effective actions can be found in the literature; the boundary terms to our knowledge are new results.

We begin with the Lipschitz-Killing curvature ${\mathcal E}_{n,m}$ and the associated boundary term ${\mathcal Q}_{n,m}$ defined in~\eqref{E:nm} and~\eqref{Qdef2} respectively. Denote the densities associated with these forms as $E_{n,m}$ and $Q_{n,m}$. The first step in evaluating the expression~\eqref{E:dWgeneral} for $\W$ is to deduce how $E_{n,m}$ and $Q_{n,m}$ change under Weyl rescalings. Starting with the metric $g_{\mu\nu}$ and performing a Weyl transformation to $\hat{g}_{\mu\nu} = e^{-2\tau}g_{\mu\nu}$, the transformed curvatures $\hat{E}_{n,m}$ and $\hat{Q}_{n,m}$ are
\begin{align}
\begin{split}
\label{E:hatEQ}
\sqrt {\hat g}~\hat E_{n,m}&=\sqrt {g}~e^{-(n-d)\tau}\left\{ E_d+{\rm D}_\mu J^\mu+(n-d)G+O(n-d)^2\right\} \ . \\ 
\sqrt {\hat \gamma}~\hat Q_{n,m}&=\sqrt {\gamma}~ e^{-(n-d)\tau}\left\{Q_d+n_\mu J^\mu +\oD_\alpha H^\alpha+(n-d)B+O(n-d)^2\right\} \,,
\end{split}
\end{align}
where it remains to determine $J^{\mu}$, $G$, $H^{\alpha}$, and $B$.
Note that, in the $n\to d$ limit,~\eqref{E:hatEQ} implies
\be
\lim_{n\to d} \left( \int_M \hat{\mathcal{E}}_{n,m} - \int_{\partial M} \hat{\mathcal{Q}}_{n,m}\right) = \lim_{n\to d}\left( \int_M \mathcal{E}_{n,m} - \int_{\partial M}\mathcal{Q}_{n,m}\right)\,,
\ee
which is just a consequence of the fact that the Euler characteristic is a topological invariant and so is invariant under Weyl rescalings. This has the practical effect that the dimensionally regulated formula~\eqref{E:dWgeneral} for $\W$ is well-defined. From~\eqref{E:hatEQ} we see that the integrand of~\eqref{E:dWgeneralv2} is 
\begin{align*}
\sqrt {\hat g}\hat E_{n,m}-\sqrt {g}E_{n,m}
&=\sqrt {g}\left\{ {\rm D}_\mu J^\mu-(n-d)\Big(\tau E_d-J^\mu\partial_{\mu}\tau-G+{\rm D}_{\mu}(\tau J^{\mu})\Big)+O(n-d)^2\right\} \ , 
\\
\sqrt {\hat \gamma}\hat Q_{n,m}-\sqrt {\gamma}Q_{n,m}&=\sqrt {\gamma}\left\{ n_\mu J^\mu +\oD_\alpha H^\alpha-(n-d)\Big(\tau Q_d+\tau (n_\mu J^\mu +\oD_\alpha H^\alpha)-B\Big) + O(n-d)^2\right\} \ . 
\end{align*}
In order to write $\W$ in as simple a way as possible, it will be useful to decompose $G$ as
\be
G = G_0 + {\rm D}_{\mu}K^{\mu}\,,
\ee
for some current $K^{\mu}$. Putting the pieces together, we find that the anomaly action $\W$ is
\begin{align}
\begin{split}
\label{E:dWgeneral2}
\W[g_{\mu\nu},e^{-2\tau}g_{\mu\nu}] = &A\left(  \int_M {\rm d}^dx \sqrt{g} \left\{ \tau E_d - J^{\mu}\partial_{\mu}\tau - G_0 \right\} \right.
\\
& \qquad \qquad \qquad \left.-  \int_{\partial M}{\rm d}^{d-1}y \sqrt{\gamma}\left\{ \tau Q_d  - H^{\alpha} \partial_{\alpha}\tau  - B + n^{\mu}K_{\mu}\right\}\right)\,.
\end{split}
\end{align}
We see that besides obtaining $B$ and $G$ defined in \eqref{E:hatEQ}, we also need to determine $J^{\mu}$, $K^{\mu}$ and $H^{\alpha}$.

\subsection{$d=4$}
 
To obtain the bulk action in $d=4$, we find that $J^{\mu}$ is
\be
J^\mu &=-8\left\{ E^{\mu\nu}\partial_\nu\tau+ (D^\mu \partial_\nu\tau) \partial^\nu\tau +(\partial^\mu\tau) (\partial\tau)^2- (\Box\tau) \partial^\mu\tau\right\}  \,,
\ee
and we find it useful to split $G$ into $G_0$ and $K^{\mu}$ as
\begin{align}
\begin{split}
K^{\mu} &= \frac{3}{2}J^{\mu} + 4 E^{\mu\nu}\partial_{\nu}\tau\, , \\
G_0 &= 4 E^{\mu\nu}(\partial_{\mu}\tau)(\partial_{\nu}\tau) -8 \Box \tau (\partial\tau)^2 + 6 (\partial\tau)^4\, \ .
\end{split}
\end{align}
We find that the boundary data $H^{\alpha}$ and $B$ are given by
\begin{align}
\begin{split}
 H^{\alpha} &=8 \left\{ \Big(K^{\alpha \beta}-\gamma^{\alpha \beta}K\Big)\partial_\beta \tau+ \tau_n \partial^\alpha \tau\right\}\,,
 \\
 B &=n^{\mu}K_{\mu}+ 4\oD_{\alpha}\left\{ \partial_{\beta}\tau\left( K^{\alpha\beta}-\gamma^{\alpha\beta} K\right)\right\} -4 \left( K^{\alpha\beta}-\gamma^{\alpha\beta} K\right)(\partial_{\alpha}\tau)(\partial_{\beta}\tau) \\ &\qquad - 8 (\oD \tau)^2\tau_n 
 - \frac{8}{3}\tau_n^3  \,,
 \end{split}
 \end{align}
where we have denoted the normal derivative of $\tau$ as $\tau_n \equiv n^{\mu}\partial_{\mu}\tau$. Substituting these expressions into the general formula~\eqref{E:dWgeneral2} for $\W$, we find the result~\eqref{Weffdfour} quoted in subsection~\ref{S:dilaton4d}.

\subsection{$d=6$}

After some tedious computation, we find that the current $J^{\mu}$ in $d=6$ for general $g_{\mu\nu}$ is given by
\be
J^{\mu}_{(6d)}=J_1^{\mu}+ J_2^{\mu}+J_3^{\mu}+J_4^{\mu}+ J_5^{\mu} \ ,
\ee
where $J^{\mu}_n$ contains $n$ powers of $\tau$, and
\be
J_1^{\mu}&=& 6 E^{(2)\mu}_{\nu} (\partial^\nu \tau) \ ,
\nn
\\ 
J_2^{\mu}&=& 48  E^{\mu}_{\nu} \Big(({\rm D}_{\rho} \partial^\nu \tau) (\partial^{\rho} \tau)-(\partial^\nu \tau) \Box \tau \Big) + 48  R^\mu{}_{\rho \nu \sigma} (\partial^\nu \tau) ({\rm D}^\sigma \partial^\rho \tau) 
\nn
 \\
&&\qquad + 48  R_{\nu \rho}  \Big((\partial^\nu \tau)  ({\rm D}^\rho \partial^\mu \tau) - ({\rm D}^\rho \partial^\nu \tau) (\partial^\mu \tau)\Big) \ ,
\nn
 \\
J_3^{\mu}&=& 48  E^{\mu}_{\nu}  (\partial^\nu \tau)  (\partial \tau)^2+ 48 (\partial^\mu \tau) (\Box \tau)^2 - 96  \Box \tau (\partial^\nu \tau)  ({\rm D}_\nu \partial^\mu \tau)
\\
&&\qquad + 96 (\partial^\nu \tau) ({\rm D}_{\rho}  \partial_\nu \tau) ({\rm D}^{\rho}  \partial^\mu \tau) - 48 ({\rm D} \partial \tau)^2 (\partial^\mu \tau)  \ , 
\nn
\\
J_4^{\mu}&=& - 144  (\partial \tau)^2 \Box \tau (\partial^\mu \tau) + 144 (\partial \tau)^2  (\partial_\rho \tau) ({\rm D}^{\rho} \partial^\mu \tau) \ , 
\nn
\\
J_5^{\mu}&=& 144 (\partial \tau)^4 (\partial^\mu \tau) \ . \nn
\ee
The quantities $E^{(2)\mu\nu}$ and $C^{\mu\nu\rho\sigma}$ are defined in~\eqref{E2def}. 

We have also computed $G$ for a general metric $g_{\mu\nu}$. We split it into $G_0$ and $K^{\mu}$ so that the bulk part of the anomaly action $\W$ matches the expression obtained in ref.~\cite{Elvang:2012st}. The resulting $K^{\mu}$ is
\be
K^{\mu}  &=& {11\over 6} J^{\mu}
- 5 E^{(2)\mu\nu} \partial_\nu \tau+ 16 E^{\mu\nu}\Big((\partial_\nu \tau) \Box \tau -  ({\rm D}^{\rho} \partial_\nu \tau) (\partial_\rho \tau) \Big) +16 C^\mu{}_{\nu\rho\sigma} ({\rm D}^{\rho} \partial^\nu \tau) (\partial^\sigma \tau) \nonumber \\
&&\qquad + 48 ({\rm D}^\mu \partial^\nu \tau) (\partial_\nu \tau) (\partial \tau)^2 + 72  (\partial \tau)^4 (\partial^\mu \tau)  - 48  (\partial \tau)^2 \Box \tau (\partial^\mu \tau) \,,
\ee
and the expression for $G_0$ is too lengthy to be worth writing here. It can be deduced by comparing the general expression for $\W$ given in~\eqref{E:dWgeneral2} with the bulk part of the anomaly action in~\eqref{Weffdsixbulk}, using the formulae for $J^{\mu}$ and $K^{\mu}$ above. 

Similarly we decompose $H^{\alpha}$ into powers of $\tau$ as
\be
H^{\alpha} = H_1^{\alpha} + H_2^{\alpha} + H_3^{\alpha} + H_4^{\alpha}\,.
\ee
The computation on the boundary becomes much more tedious. We have computed $B$ in general
but its expression is too lengthy to present here. We have not yet succeeded in finding the current $H^{\alpha}$ when for a general metric $g_{\mu\nu}$. When $\hat{g}_{\mu\nu}$ is conformally flat, $\hat{g}_{\mu\nu}=e^{-2\tau}\delta_{\mu\nu}$, we find
\begin{align}
\begin{split}
H^\alpha_{1}&= 48 P^\alpha_\beta \partial^{\beta} \tau  + 6 Q_4[\delta_{\mu\nu}] \partial^\alpha \tau \ ,
\\
H^\alpha_{2} &= 48 K^\alpha_\beta (\partial^\beta \tau)  \oBox \tau-48 K^\alpha_\beta (\oD_\gamma \partial^\beta \tau) (\partial^\gamma \tau)-48 K  (\partial^\alpha \tau) \oBox \tau 
\\
&\qquad +48 K^\beta_\gamma (\oD^\gamma \partial_\beta \tau) (\partial^\alpha \tau)+48 K (\partial_\beta \tau) (\oD^\alpha \partial^\beta \tau) -48 K^\beta_\gamma (\oD^\alpha \partial_\beta \tau) (\partial^\gamma \tau) \ , 
\\
H^\alpha_{3} &=- 48 K^\alpha_{\beta}  (\partial^{\beta}\tau)  (\oD \tau)^2  + 48 K (\oD \tau)^2  (\partial^\alpha \tau)+ 48  K  \tau_n^2  (\partial^\alpha \tau)- 48 \tau_n^2 K^\alpha_{\beta}  (\partial^{\beta} \tau) 
 \\
 &\qquad + 96  \tau_n \oBox \tau (\partial^\alpha \tau)- 96  \tau_n (\oD^\alpha \partial_{\beta}\tau) (\partial^{\beta}\tau) \ , 
 \\
H^\alpha_{4}&= - 144   \tau_n (\oD \tau)^2  \partial^\alpha \tau- 48  \tau_n^3 (\partial^\alpha \tau)\,, 
\end{split}
\end{align}
where we defined $P^{\alpha \beta}$ in~\eqref{Pabdef}. Using the expressions present above and the general expression for the boundary term of $\W$ in~\eqref{E:dWgeneral2}, we obtain the explicit form in $d=6$ given in \eqref{6dresult}.

\section{Holographic Calculation}
\label{app:holo}

In this appendix, we study $d$-dimensional conformal field theories with a dual gravitational description via the AdS/CFT correspondence. 
We then use the correspondence to compute the thermodynamics of these conformal field theories when they live on a hyperbolic space $H^{d-1}$ with radius of curvature $\ell$ at temperature $T$. In the special case $T = 1/(2 \pi \ell)$, we will be able to compare with the previous anomaly calculations.

Much of the following calculation can be found already in ref.\ \cite{Emparan:1999gf} and~\cite{Casini:2011kv}. In particular, the expression for the thermal entropy on $H^{d-1}$ at temperature $T=1/(2\pi \ell)$ in terms of $a$ is given in section 3 of~\cite{Casini:2011kv}. Our new result is the thermal partition function on hyperbolic space at any temperature and in any $d$.

We start with the usual bulk plus Gibbons-Hawking plus counterterm action for these holographic calculations (see for example ref.\ \cite{Emparan:1999pm}):
\begin{align}
\begin{split}
S &= S_{\rm bulk} + S_{\rm surf} + S_{\rm ct} \ , \\
S_{\rm bulk} &= - \frac{1}{2 \kappa^2} \int_{\mathcal M} \d^{d+1} X \, \sqrt{-G} \left\{ {\mathcal R} + \frac{d(d-1)}{L^2} \right\} \ , \\
S_{\rm GH} &= - \frac{1}{\kappa^2} \int_{\partial{\mathcal M}} \d^{d} x \, \sqrt{-g} K \ , \\
S_{\rm CT} &= \frac{1}{2 \kappa^2} \int_{\partial {\mathcal M}} \d^d x \sqrt{-g} \biggl[ \frac{2(d-1)}{L} + \frac{L}{d-2} R  \\
&\qquad \qquad \qquad \qquad \qquad +\frac{L^3}{(d-4)(d-2)^2} \left( R^{\mu\nu} R_{\mu\nu} - \frac{d}{4(d-1)} R^2 \right) + \ldots \biggr] \ .
\end{split}
\end{align}
We denote the bulk metric as $G$, bulk coordinates as $X$, and $\mathcal{R}$ is the bulk scalar curvature. The bulk spacetime $\mathcal{M}$ is asymptotically AdS, and so the on-shell Einstein-Hilbert action $S_{\rm bulk}$ diverges owing to the infinite volume ``near'' the AdS boundary. To compute thermodynamic quantities, we must holographically renormalize the bulk gravity. In the usual way, we introduce a ``cutoff surface'' $\partial\mathcal{M}$ near the AdS boundary; the induced metric on the cutoff surface is $g$, coordinates on it are denoted as $x^{\mu}$, and $R^{\mu}{}_{\nu\rho\sigma}$ refers to the Riemann tensor constructed from $g$. We introduce the Gibbons-Hawking term on this cutoff surface, along with various counterterms $S_{\rm CT}$, and ultimately take the limit where we send the cutoff surface to the AdS boundary. The counterterms are tuned so that this limit exists.

To obtain the thermodynamic partition function $W_H = - \ln Z_H$ on hyperbolic space, we first identify the gravitational solution dual to the thermal state on hyperbolic space, namely the AdS-black hole with hyperbolic boundary. We then Wick rotate the bulk spacetime to Euclidean signature and compute the on-shell, holographically renormalized, Euclidean action. 

The AdS-black hole metric with hyperbolic boundary is a solution to the equations of motion:
\begin{equation}
\label{E:hyperbolicBH}
G = - \left( \frac{r^2}{L^2} f(r) - 1 \right) \frac{L^2}{\ell^2} \d t^2 + r^2 (\d u^2 + \sinh^2 u \, \d \Omega_{d-2} ) + \frac{\d r^2}{\frac{r^2}{L^2} f(r) - 1} \,,
\end{equation}
with $f(r) = 1 - m/ r^d$ where $m$ is an integration constant related to the temperature. The constant $m$ can be expressed in terms of the horizon radius $r_h$: 
$
m = (r_h^2 - L^2)   r_h^{d-2}
$.
In terms of the horizon radius, the temperature is 
\be
T = \frac{1}{\beta} = \frac{  r_h^2 d - (d-2)L^2}{4 \pi L \ell r_h} \ ,
\ee
which can be inverted to give the horizon radius as a function of $\beta$:
\[
\frac{r_h}{L} d = \frac{2 \pi \ell}{\beta} + \sqrt{d(d-2)+\left( \frac{2 \pi \ell}{\beta}\right)^2 } \ .
\]
Note that at  
$m=0$, the metric becomes that of pure AdS with a hyperbolic slicing, and the horizon radius is the same as the radius $L$ of curvature of AdS. The temperature at this point is $T = 1/(2\pi \ell)$, and the black hole is ``topological'' in the sense that it is simply a causal horizon.

The most direct way to check the entanglement entropy calculation is to compute the area of the black hole horizon and use the Bekenstein-Hawking area law for black hole entropy. One finds straightforwardly that 
\be
S_{\rm BH} = \frac{2 \pi r_h^{d-1} }{\ell^{d-1} \kappa^2} \Vol(H^{d-1}) \,,
\label{Shologeneral}
\ee
where the hyperbolic space has radius of curvature $\ell$. This entropy diverges for the simple reason that hyperbolic space has infinite volume, in the same way that the total entropy in flat space diverges. However, unlike in flat space, we may appropriately regulate the volume of $H^{d-1}$ and thereby identify a universal logarithmic term in $\Vol(H^{d-1})$ as in~\eqref{E:hyperbolicVolume}.
To check the calculation of the entanglement entropy across a sphere in flat space, we work with the ``topological'' black hole at $T = 1/(2\pi \ell)$ with horizon radius $r_h=L$.

To compare the holographic entropy result (\ref{Shologeneral}) with field theory, we need an expression relating $a$
to the gravitational coupling constant $\kappa$ in general dimension:
\be
\label{atokappa}
a =\frac{1}{2} \Vol(S^{d-1}) \frac{L^{d-1}}{\kappa^2} \ . 
\ee
This relation is consistent with the holographic Weyl anomaly computed in $d=2$, 4 and 6 dimensions in ref.\ \cite{Henningson:1998gx}. In general $d$, this relation can be extracted from ref.~\cite{Imbimbo:1999bj}.\footnote{%
It is straightforward to derive eq.\ (\ref{atokappa}) by placing the field theory on an $S^d$, computing the Euclidean partition function and using the relation
\[
W_{S^d} = - \ln Z_{S^d} = (-1)^{d/2} 4a \ln (\mu \ell) + \hdots\,,
\]
where $\mu$ is an energy scale introduced in the course of defining the theory. The ``sphere free energy'' $W_{S^d}$ is equal to the holographically renormalized, on-shell action $S$ evaluated on the asymptotically hyperbolic metric with $S^d$ boundary,
\[
G = r^2 \ell^2 \d\Omega_d^2 +L^2 \frac{\d r^2}{r^2+L^2/\ell^2}\,.
\]
The logarithmic ambiguity in $S$ arises purely from a logarithmic divergence in the on-shell bulk action $S_{\rm bulk}$ at large $r$, and using some of the same steps we employ below to compute the partition function on $S^1 \times H^{d-1}$, we find
eq. (\ref{atokappa}).}
 As we did in the previous section, we now extract the logarithmic contribution to $\Vol(H^{d-1})$ and use the formula~\eqref{atokappa} for $a$ to obtain
\be
S_{\rm BH} = S_E = \ldots + (-1)^{d/2} 4a  \ln \frac{\delta}{\ell} + \ldots
\ee
in agreement with the universal result~\eqref{SEball}. This holographic result was also obtained in~\cite{Casini:2011kv} although their result is stronger as it allows for higher derivative curvature corrections to the gravity action.

We are also interested in looking at the partition function $W_H$ which can be equated holographically to the on-shell value of the gravity action on the Euclidean version of~\eqref{E:hyperbolicBH}. Einstein's equations imply that the Lagrangian density evaluates to
\be
{\mathcal R} + \frac{d(d-1)}{L^2} = - \frac{2 d}{L^2} \ 
\ee
on shell.
To avoid a lengthy discussion of counter-terms, we note that because the time direction in the boundary is flat, the counterterms can  depend on $r_h$ only through the metric determinant $\sqrt{g}$.  It is therefore convenient to divide out a factor of $\sqrt{-g_{tt}}$ from on-shell quantities. The bulk and Gibbons-Hawking actions evaluate to 
\begin{align}
\begin{split}
\frac{S_{\rm bulk} + S_{\rm GH}}{ \frac{\ell}{r} \sqrt{-g_{tt}}} = &\left[ (d-1) \left( \frac{r}{L} \right)^d \sqrt{ 1- \left( \frac{L}{r} \right)^2}
- \frac{1}{2} \left( \frac{r_h}{L} \right)^d - \frac{1}{2} \left( \frac{r_h}{L} \right)^{d-2} + O(r^{-2}) \right] 
\\
& \qquad \qquad \qquad \times \frac{ \beta L^{d-1} \Vol(H^{d-1})}{\ell \kappa^2}\ .
\end{split}
\end{align}
The counterterms should be whatever they need to be to cancel the divergent factors coming from the square root. By dimensional analysis, a counterterm with $2n$ derivatives of the boundary metric will cancel a divergence at $O(r^{d-2n})$. In a minimal counterterm prescription where we add no finite terms with $d$ derivatives, e.g. $(R_{\mu\nu\rho\sigma})^{d/2}$, expanding out the square root, the on-shell action is
\be
W_H = -\left[ \frac{(d-1)!}{2^{d-1} \left( \frac{d}{2} \right)! \left( \frac{d-2}{2} \right)! } - \frac{1}{2} \left( \frac{r_h}{L} \right)^d - \frac{1}{2} \left( \frac{r_h}{L} \right)^{d-2} \right]  \frac{ \beta L^{d-1} \Vol(H^{d-1})}{\ell^d \kappa^2} \ .
\ee
We have the partition function as a function of $\beta$ and $\ell$ and not just in the ``topological'' limit $\beta = 2 \pi \ell$.

It is straightforward to verify the black hole entropy calculation above using standard thermodynamic identities. We can compute the thermal energy from the effective action by taking a $\beta$ derivative:
\be
\langle H \rangle = -\frac{\partial W_H}{\partial \beta} \ .
\ee
The black hole entropy is then $S_{\rm BH} = \beta \langle H \rangle - W_H$, in agreement with the event horizon area~\eqref{Shologeneral}. Note that the energy and $W_H$ itself are ambiguous quantities. The first term in $W_H$ can be altered if we decide to add a local counterterm like $(R_{\mu\nu\rho\sigma})^{d/2}$. Because the first term is linear in $\beta$, the energy suffers a similar ambiguity, but this scheme-dependence drops out of the black hole entropy.  

Because not all field theories
have classical gravity duals, this partition function  
will not hold generally, but we can compare with the other parts of the paper when $\beta = 2 \pi \ell$. In the ``topological'' case, making use of the expression~\eqref{atokappa} for $a$, we see that $W_H$ agrees with the general CFT result~\eqref{Wholo}.
Interestingly, the derivative of the $r_h$ dependent terms of $W_H$ with respect to $\beta$ vanishes at $\beta = 2 \pi \ell$.  Thus the entire contribution to the energy comes from the first (regulator dependent) piece linear in $\beta$ when $\beta = 2 \pi \ell$: 
\be
\langle H \rangle = -\frac{\partial W_H}{\partial\beta} = 2a \frac{ \Gamma(d)}{(4 \pi \ell^2)^{d/2} \left( \frac{d}{2} \right)!}
\Vol(H^{d-1}) \ .
\ee
Note this result agrees with the general CFT calculation (\ref{CFTH}) as well.

A peculiar observation about this holographic thermal partition function is that
\be
\label{holosecderiv}
\left. \frac{\partial^2}{\partial \ell \, \partial \beta} \left( \frac{W_H}{\ell \beta} \right)\right|_{\beta = 2 \pi \ell} = 0 \ .
\ee
Note that $W_H = f(2 \pi \ell / \beta)$ is essentially a function of one variable, the ratio $2 \pi \ell / \beta$. It follows that $\beta \partial_\beta W_H = - \ell \partial_\ell W_H$. As $\partial_\beta W_H$ is proportional to the energy while $\partial_\ell W_H$ is proportional to a trace of the stress tensor over the $H^{d-1}$ directions, the fact that $W_H$ depends on $\ell/ \beta$ encodes the fact that the integral of the trace of the stress tensor vanishes. The relation~\eqref{holosecderiv} is a stronger statement, which naively relates integrals of the two-point function of the stress tensor. Perhaps it follows from the form of the two-point function of the stress tensor on $S^1\times H^{d-1}$ at $T = 1/(2\pi \ell)$, which is determined by conformal symmetry.


\end{document}